\documentclass[12pt]{article}	
\usepackage{amsmath, amssymb,amsmath,amsthm,palatino} 					
\usepackage{bm}

\usepackage{amsmath, amssymb,amsmath,amsthm} 
\usepackage{bm}														
\usepackage[usenames,dvipsnames]{color}								
\usepackage{setspace}	

\usepackage{geometry}												
\usepackage[sc]{mathpazo}

\usepackage[T1]{fontenc}
\usepackage[utf8]{inputenc}
\usepackage[english]{babel}

\usepackage{graphicx}													

\usepackage{url}																

\usepackage[T1]{fontenc}

\usepackage[plainpages=false]{hyperref}					
		
\usepackage[usenames,dvipsnames,svgnames,table]{xcolor}
\hypersetup{
 colorlinks=true,
 citecolor=Maroon,	
 linkcolor=Blue,
 urlcolor=Maroon}																						

\usepackage{pst-all}                            

\usepackage{nicefrac}														
\usepackage{booktabs}
\usepackage{float}

\newcommand{\vect}[1]{\mathbf{#1}}

\newtheorem{Assumption}{Assumption}
\newtheorem{Theorem}{Theorem}

\newtheorem{Corollary}{Corollary}

\newtheorem{Lemma}{Lemma}
\newtheorem*{Proof*}{Proof}

\newtheorem{Remark}{Remark}

\definecolor{snow}{rgb}{1.000,0.980,0.980}
\definecolor{snow1}{rgb}{1.000,0.980,0.980}
\definecolor{snow2}{rgb}{0.933,0.914,0.914}
\definecolor{snow3}{rgb}{0.804,0.788,0.788}
\definecolor{snow4}{rgb}{0.545,0.537,0.537}
\definecolor{GhostWhite}{rgb}{0.973,0.973,1.000}
\definecolor{WhiteSmoke}{rgb}{0.961,0.961,0.961}
\definecolor{gainsboro}{rgb}{0.863,0.863,0.863}
\definecolor{FloralWhite}{rgb}{1.000,0.980,0.941}
\definecolor{OldLace}{rgb}{0.992,0.961,0.902}
\definecolor{linen}{rgb}{0.980,0.941,0.902}
\definecolor{AntiqueWhite}{rgb}{0.980,0.922,0.843}
\definecolor{PapayaWhip}{rgb}{1.000,0.937,0.835}
\definecolor{BlanchedAlmond}{rgb}{1.000,0.922,0.804}
\definecolor{bisque}{rgb}{1.000,0.894,0.769}
\definecolor{PeachPuff}{rgb}{1.000,0.855,0.725}
\definecolor{NavajoWhite}{rgb}{1.000,0.871,0.678}
\definecolor{moccasin}{rgb}{1.000,0.894,0.710}
\definecolor{cornsilk}{rgb}{1.000,0.973,0.863}
\definecolor{ivory}{rgb}{1.000,1.000,0.941}
\definecolor{LemonChiffon}{rgb}{1.000,0.980,0.804}
\definecolor{seashell}{rgb}{1.000,0.961,0.933}
\definecolor{honeydew}{rgb}{0.941,1.000,0.941}
\definecolor{MintCream}{rgb}{0.961,1.000,0.980}
\definecolor{azure}{rgb}{0.941,1.000,1.000}
\definecolor{AliceBlue}{rgb}{0.941,0.973,1.000}
\definecolor{lavender}{rgb}{0.902,0.902,0.980}
\definecolor{LavenderBlush}{rgb}{1.000,0.941,0.961}
\definecolor{MistyRose}{rgb}{1.000,0.894,0.882}
\definecolor{white}{rgb}{1.000,1.000,1.000}
\definecolor{black}{rgb}{0.000,0.000,0.000}
\definecolor{DarkSlateGray}{rgb}{0.184,0.310,0.310}
\definecolor{DimGray}{rgb}{0.412,0.412,0.412}
\definecolor{SlateGray}{rgb}{0.439,0.502,0.565}
\definecolor{LightSlateGray}{rgb}{0.467,0.533,0.600}
\definecolor{gray}{rgb}{0.745,0.745,0.745}
\definecolor{LightGray}{rgb}{0.827,0.827,0.827}
\definecolor{MidnightBlue}{rgb}{0.098,0.098,0.439}
\definecolor{navy}{rgb}{0.000,0.000,0.502}
\definecolor{NavyBlue}{rgb}{0.000,0.000,0.502}
\definecolor{CornflowerBlue}{rgb}{0.392,0.584,0.929}
\definecolor{DarkSlateBlue}{rgb}{0.282,0.239,0.545}
\definecolor{SlateBlue}{rgb}{0.416,0.353,0.804}
\definecolor{MediumSlateBlue}{rgb}{0.482,0.408,0.933}
\definecolor{LightSlateBlue}{rgb}{0.518,0.439,1.000}
\definecolor{MediumBlue}{rgb}{0.000,0.000,0.804}
\definecolor{RoyalBlue}{rgb}{0.255,0.412,0.882}
\definecolor{blue}{rgb}{0.000,0.000,1.000}
\definecolor{DodgerBlue}{rgb}{0.118,0.565,1.000}
\definecolor{DeepSkyBlue}{rgb}{0.000,0.749,1.000}
\definecolor{SkyBlue}{rgb}{0.529,0.808,0.922}
\definecolor{LightSkyBlue}{rgb}{0.529,0.808,0.980}
\definecolor{SteelBlue}{rgb}{0.275,0.510,0.706}
\definecolor{LightSteelBlue}{rgb}{0.690,0.769,0.871}
\definecolor{LightBlue}{rgb}{0.678,0.847,0.902}
\definecolor{PowderBlue}{rgb}{0.690,0.878,0.902}
\definecolor{PaleTurquoise}{rgb}{0.686,0.933,0.933}
\definecolor{DarkTurquoise}{rgb}{0.000,0.808,0.820}
\definecolor{MediumTurquoise}{rgb}{0.282,0.820,0.800}
\definecolor{turquoise}{rgb}{0.251,0.878,0.816}
\definecolor{cyan}{rgb}{0.000,1.000,1.000}
\definecolor{LightCyan}{rgb}{0.878,1.000,1.000}
\definecolor{CadetBlue}{rgb}{0.373,0.620,0.627}
\definecolor{MediumAquamarine}{rgb}{0.400,0.804,0.667}
\definecolor{aquamarine}{rgb}{0.498,1.000,0.831}
\definecolor{DarkGreen}{rgb}{0.000,0.392,0.000}
\definecolor{DarkOliveGreen}{rgb}{0.333,0.420,0.184}
\definecolor{DarkSeaGreen}{rgb}{0.561,0.737,0.561}
\definecolor{SeaGreen}{rgb}{0.180,0.545,0.341}
\definecolor{MediumSeaGreen}{rgb}{0.235,0.702,0.443}
\definecolor{LightSeaGreen}{rgb}{0.125,0.698,0.667}
\definecolor{PaleGreen}{rgb}{0.596,0.984,0.596}
\definecolor{SpringGreen}{rgb}{0.000,1.000,0.498}
\definecolor{LawnGreen}{rgb}{0.486,0.988,0.000}
\definecolor{green}{rgb}{0.000,1.000,0.000}
\definecolor{chartreuse}{rgb}{0.498,1.000,0.000}
\definecolor{MediumSpringGreen}{rgb}{0.000,0.980,0.604}
\definecolor{GreenYellow}{rgb}{0.678,1.000,0.184}
\definecolor{LimeGreen}{rgb}{0.196,0.804,0.196}
\definecolor{YellowGreen}{rgb}{0.604,0.804,0.196}
\definecolor{ForestGreen}{rgb}{0.133,0.545,0.133}
\definecolor{OliveDrab}{rgb}{0.420,0.557,0.137}
\definecolor{DarkKhaki}{rgb}{0.741,0.718,0.420}
\definecolor{khaki}{rgb}{0.941,0.902,0.549}
\definecolor{PaleGoldenrod}{rgb}{0.933,0.910,0.667}
\definecolor{LightGoldenrodYellow}{rgb}{0.980,0.980,0.824}
\definecolor{LightYellow}{rgb}{1.000,1.000,0.878}
\definecolor{yellow}{rgb}{1.000,1.000,0.000}
\definecolor{gold}{rgb}{1.000,0.843,0.000}
\definecolor{LightGoldenrod}{rgb}{0.933,0.867,0.510}
\definecolor{goldenrod}{rgb}{0.855,0.647,0.125}
\definecolor{DarkGoldenrod}{rgb}{0.722,0.525,0.043}
\definecolor{RosyBrown}{rgb}{0.737,0.561,0.561}
\definecolor{IndianRed}{rgb}{0.804,0.361,0.361}
\definecolor{SaddleBrown}{rgb}{0.545,0.271,0.075}
\definecolor{sienna}{rgb}{0.627,0.322,0.176}
\definecolor{peru}{rgb}{0.804,0.522,0.247}
\definecolor{burlywood}{rgb}{0.871,0.722,0.529}
\definecolor{beige}{rgb}{0.961,0.961,0.863}
\definecolor{wheat}{rgb}{0.961,0.871,0.702}
\definecolor{SandyBrown}{rgb}{0.957,0.643,0.376}
\definecolor{tan}{rgb}{0.824,0.706,0.549}
\definecolor{chocolate}{rgb}{0.824,0.412,0.118}
\definecolor{firebrick}{rgb}{0.698,0.133,0.133}
\definecolor{brown}{rgb}{0.647,0.165,0.165}
\definecolor{DarkSalmon}{rgb}{0.914,0.588,0.478}
\definecolor{salmon}{rgb}{0.980,0.502,0.447}
\definecolor{LightSalmon}{rgb}{1.000,0.627,0.478}
\definecolor{orange}{rgb}{1.000,0.647,0.000}
\definecolor{DarkOrange}{rgb}{1.000,0.549,0.000}
\definecolor{coral}{rgb}{1.000,0.498,0.314}
\definecolor{LightCoral}{rgb}{0.941,0.502,0.502}
\definecolor{tomato}{rgb}{1.000,0.388,0.278}
\definecolor{OrangeRed}{rgb}{1.000,0.271,0.000}
\definecolor{red}{rgb}{1.000,0.000,0.000}
\definecolor{HotPink}{rgb}{1.000,0.412,0.706}
\definecolor{DeepPink}{rgb}{1.000,0.078,0.576}
\definecolor{pink}{rgb}{1.000,0.753,0.796}
\definecolor{LightPink}{rgb}{1.000,0.714,0.757}
\definecolor{PaleVioletRed}{rgb}{0.859,0.439,0.576}
\definecolor{maroon}{rgb}{0.690,0.188,0.376}
\definecolor{MediumVioletRed}{rgb}{0.780,0.082,0.522}
\definecolor{VioletRed}{rgb}{0.816,0.125,0.565}
\definecolor{magenta}{rgb}{1.000,0.000,1.000}
\definecolor{violet}{rgb}{0.933,0.510,0.933}
\definecolor{plum}{rgb}{0.867,0.627,0.867}
\definecolor{orchid}{rgb}{0.855,0.439,0.839}
\definecolor{MediumOrchid}{rgb}{0.729,0.333,0.827}
\definecolor{DarkOrchid}{rgb}{0.600,0.196,0.800}
\definecolor{DarkViolet}{rgb}{0.580,0.000,0.827}
\definecolor{BlueViolet}{rgb}{0.541,0.169,0.886}
\definecolor{purple}{rgb}{0.627,0.125,0.941}
\definecolor{MediumPurple}{rgb}{0.576,0.439,0.859}
\definecolor{thistle}{rgb}{0.847,0.749,0.847}
\definecolor{seashell1}{rgb}{1.000,0.961,0.933}
\definecolor{seashell2}{rgb}{0.933,0.898,0.871}
\definecolor{seashell3}{rgb}{0.804,0.773,0.749}
\definecolor{seashell4}{rgb}{0.545,0.525,0.510}
\definecolor{AntiqueWhite1}{rgb}{1.000,0.937,0.859}
\definecolor{AntiqueWhite2}{rgb}{0.933,0.875,0.800}
\definecolor{AntiqueWhite3}{rgb}{0.804,0.753,0.690}
\definecolor{AntiqueWhite4}{rgb}{0.545,0.514,0.471}
\definecolor{bisque1}{rgb}{1.000,0.894,0.769}
\definecolor{bisque2}{rgb}{0.933,0.835,0.718}
\definecolor{bisque3}{rgb}{0.804,0.718,0.620}
\definecolor{bisque4}{rgb}{0.545,0.490,0.420}
\definecolor{PeachPuff1}{rgb}{1.000,0.855,0.725}
\definecolor{PeachPuff2}{rgb}{0.933,0.796,0.678}
\definecolor{PeachPuff3}{rgb}{0.804,0.686,0.584}
\definecolor{PeachPuff4}{rgb}{0.545,0.467,0.396}
\definecolor{NavajoWhite1}{rgb}{1.000,0.871,0.678}
\definecolor{NavajoWhite2}{rgb}{0.933,0.812,0.631}
\definecolor{NavajoWhite3}{rgb}{0.804,0.702,0.545}
\definecolor{NavajoWhite4}{rgb}{0.545,0.475,0.369}
\definecolor{LemonChiffon1}{rgb}{1.000,0.980,0.804}
\definecolor{LemonChiffon2}{rgb}{0.933,0.914,0.749}
\definecolor{LemonChiffon3}{rgb}{0.804,0.788,0.647}
\definecolor{LemonChiffon4}{rgb}{0.545,0.537,0.439}
\definecolor{cornsilk1}{rgb}{1.000,0.973,0.863}
\definecolor{cornsilk2}{rgb}{0.933,0.910,0.804}
\definecolor{cornsilk3}{rgb}{0.804,0.784,0.694}
\definecolor{cornsilk4}{rgb}{0.545,0.533,0.471}
\definecolor{ivory1}{rgb}{1.000,1.000,0.941}
\definecolor{ivory2}{rgb}{0.933,0.933,0.878}
\definecolor{ivory3}{rgb}{0.804,0.804,0.757}
\definecolor{ivory4}{rgb}{0.545,0.545,0.514}
\definecolor{honeydew1}{rgb}{0.941,1.000,0.941}
\definecolor{honeydew2}{rgb}{0.878,0.933,0.878}
\definecolor{honeydew3}{rgb}{0.757,0.804,0.757}
\definecolor{honeydew4}{rgb}{0.514,0.545,0.514}
\definecolor{LavenderBlush1}{rgb}{1.000,0.941,0.961}
\definecolor{LavenderBlush2}{rgb}{0.933,0.878,0.898}
\definecolor{LavenderBlush3}{rgb}{0.804,0.757,0.773}
\definecolor{LavenderBlush4}{rgb}{0.545,0.514,0.525}
\definecolor{MistyRose1}{rgb}{1.000,0.894,0.882}
\definecolor{MistyRose2}{rgb}{0.933,0.835,0.824}
\definecolor{MistyRose3}{rgb}{0.804,0.718,0.710}
\definecolor{MistyRose4}{rgb}{0.545,0.490,0.482}
\definecolor{azure1}{rgb}{0.941,1.000,1.000}
\definecolor{azure2}{rgb}{0.878,0.933,0.933}
\definecolor{azure3}{rgb}{0.757,0.804,0.804}
\definecolor{azure4}{rgb}{0.514,0.545,0.545}
\definecolor{SlateBlue1}{rgb}{0.514,0.435,1.000}
\definecolor{SlateBlue2}{rgb}{0.478,0.404,0.933}
\definecolor{SlateBlue3}{rgb}{0.412,0.349,0.804}
\definecolor{SlateBlue4}{rgb}{0.278,0.235,0.545}
\definecolor{RoyalBlue1}{rgb}{0.282,0.463,1.000}
\definecolor{RoyalBlue2}{rgb}{0.263,0.431,0.933}
\definecolor{RoyalBlue3}{rgb}{0.227,0.373,0.804}
\definecolor{RoyalBlue4}{rgb}{0.153,0.251,0.545}
\definecolor{blue1}{rgb}{0.000,0.000,1.000}
\definecolor{blue2}{rgb}{0.000,0.000,0.933}
\definecolor{blue3}{rgb}{0.000,0.000,0.804}
\definecolor{blue4}{rgb}{0.000,0.000,0.545}
\definecolor{DodgerBlue1}{rgb}{0.118,0.565,1.000}
\definecolor{DodgerBlue2}{rgb}{0.110,0.525,0.933}
\definecolor{DodgerBlue3}{rgb}{0.094,0.455,0.804}
\definecolor{DodgerBlue4}{rgb}{0.063,0.306,0.545}
\definecolor{SteelBlue1}{rgb}{0.388,0.722,1.000}
\definecolor{SteelBlue2}{rgb}{0.361,0.675,0.933}
\definecolor{SteelBlue3}{rgb}{0.310,0.580,0.804}
\definecolor{SteelBlue4}{rgb}{0.212,0.392,0.545}
\definecolor{DeepSkyBlue1}{rgb}{0.000,0.749,1.000}
\definecolor{DeepSkyBlue2}{rgb}{0.000,0.698,0.933}
\definecolor{DeepSkyBlue3}{rgb}{0.000,0.604,0.804}
\definecolor{DeepSkyBlue4}{rgb}{0.000,0.408,0.545}
\definecolor{SkyBlue1}{rgb}{0.529,0.808,1.000}
\definecolor{SkyBlue2}{rgb}{0.494,0.753,0.933}
\definecolor{SkyBlue3}{rgb}{0.424,0.651,0.804}
\definecolor{SkyBlue4}{rgb}{0.290,0.439,0.545}
\definecolor{LightSkyBlue1}{rgb}{0.690,0.886,1.000}
\definecolor{LightSkyBlue2}{rgb}{0.643,0.827,0.933}
\definecolor{LightSkyBlue3}{rgb}{0.553,0.714,0.804}
\definecolor{LightSkyBlue4}{rgb}{0.376,0.482,0.545}
\definecolor{SlateGray1}{rgb}{0.776,0.886,1.000}
\definecolor{SlateGray2}{rgb}{0.725,0.827,0.933}
\definecolor{SlateGray3}{rgb}{0.624,0.714,0.804}
\definecolor{SlateGray4}{rgb}{0.424,0.482,0.545}
\definecolor{LightSteelBlue1}{rgb}{0.792,0.882,1.000}
\definecolor{LightSteelBlue2}{rgb}{0.737,0.824,0.933}
\definecolor{LightSteelBlue3}{rgb}{0.635,0.710,0.804}
\definecolor{LightSteelBlue4}{rgb}{0.431,0.482,0.545}
\definecolor{LightBlue1}{rgb}{0.749,0.937,1.000}
\definecolor{LightBlue2}{rgb}{0.698,0.875,0.933}
\definecolor{LightBlue3}{rgb}{0.604,0.753,0.804}
\definecolor{LightBlue4}{rgb}{0.408,0.514,0.545}
\definecolor{LightCyan1}{rgb}{0.878,1.000,1.000}
\definecolor{LightCyan2}{rgb}{0.820,0.933,0.933}
\definecolor{LightCyan3}{rgb}{0.706,0.804,0.804}
\definecolor{LightCyan4}{rgb}{0.478,0.545,0.545}
\definecolor{PaleTurquoise1}{rgb}{0.733,1.000,1.000}
\definecolor{PaleTurquoise2}{rgb}{0.682,0.933,0.933}
\definecolor{PaleTurquoise3}{rgb}{0.588,0.804,0.804}
\definecolor{PaleTurquoise4}{rgb}{0.400,0.545,0.545}
\definecolor{CadetBlue1}{rgb}{0.596,0.961,1.000}
\definecolor{CadetBlue2}{rgb}{0.557,0.898,0.933}
\definecolor{CadetBlue3}{rgb}{0.478,0.773,0.804}
\definecolor{CadetBlue4}{rgb}{0.325,0.525,0.545}
\definecolor{turquoise1}{rgb}{0.000,0.961,1.000}
\definecolor{turquoise2}{rgb}{0.000,0.898,0.933}
\definecolor{turquoise3}{rgb}{0.000,0.773,0.804}
\definecolor{turquoise4}{rgb}{0.000,0.525,0.545}
\definecolor{cyan1}{rgb}{0.000,1.000,1.000}
\definecolor{cyan2}{rgb}{0.000,0.933,0.933}
\definecolor{cyan3}{rgb}{0.000,0.804,0.804}
\definecolor{cyan4}{rgb}{0.000,0.545,0.545}
\definecolor{DarkSlateGray1}{rgb}{0.592,1.000,1.000}
\definecolor{DarkSlateGray2}{rgb}{0.553,0.933,0.933}
\definecolor{DarkSlateGray3}{rgb}{0.475,0.804,0.804}
\definecolor{DarkSlateGray4}{rgb}{0.322,0.545,0.545}
\definecolor{aquamarine1}{rgb}{0.498,1.000,0.831}
\definecolor{aquamarine2}{rgb}{0.463,0.933,0.776}
\definecolor{aquamarine3}{rgb}{0.400,0.804,0.667}
\definecolor{aquamarine4}{rgb}{0.271,0.545,0.455}
\definecolor{DarkSeaGreen1}{rgb}{0.757,1.000,0.757}
\definecolor{DarkSeaGreen2}{rgb}{0.706,0.933,0.706}
\definecolor{DarkSeaGreen3}{rgb}{0.608,0.804,0.608}
\definecolor{DarkSeaGreen4}{rgb}{0.412,0.545,0.412}
\definecolor{SeaGreen1}{rgb}{0.329,1.000,0.624}
\definecolor{SeaGreen2}{rgb}{0.306,0.933,0.580}
\definecolor{SeaGreen3}{rgb}{0.263,0.804,0.502}
\definecolor{SeaGreen4}{rgb}{0.180,0.545,0.341}
\definecolor{PaleGreen1}{rgb}{0.604,1.000,0.604}
\definecolor{PaleGreen2}{rgb}{0.565,0.933,0.565}
\definecolor{PaleGreen3}{rgb}{0.486,0.804,0.486}
\definecolor{PaleGreen4}{rgb}{0.329,0.545,0.329}
\definecolor{SpringGreen1}{rgb}{0.000,1.000,0.498}
\definecolor{SpringGreen2}{rgb}{0.000,0.933,0.463}
\definecolor{SpringGreen3}{rgb}{0.000,0.804,0.400}
\definecolor{SpringGreen4}{rgb}{0.000,0.545,0.271}
\definecolor{green1}{rgb}{0.000,1.000,0.000}
\definecolor{green2}{rgb}{0.000,0.933,0.000}
\definecolor{green3}{rgb}{0.000,0.804,0.000}
\definecolor{green4}{rgb}{0.000,0.545,0.000}
\definecolor{chartreuse1}{rgb}{0.498,1.000,0.000}
\definecolor{chartreuse2}{rgb}{0.463,0.933,0.000}
\definecolor{chartreuse3}{rgb}{0.400,0.804,0.000}
\definecolor{chartreuse4}{rgb}{0.271,0.545,0.000}
\definecolor{OliveDrab1}{rgb}{0.753,1.000,0.243}
\definecolor{OliveDrab2}{rgb}{0.702,0.933,0.227}
\definecolor{OliveDrab3}{rgb}{0.604,0.804,0.196}
\definecolor{OliveDrab4}{rgb}{0.412,0.545,0.133}
\definecolor{DarkOliveGreen1}{rgb}{0.792,1.000,0.439}
\definecolor{DarkOliveGreen2}{rgb}{0.737,0.933,0.408}
\definecolor{DarkOliveGreen3}{rgb}{0.635,0.804,0.353}
\definecolor{DarkOliveGreen4}{rgb}{0.431,0.545,0.239}
\definecolor{khaki1}{rgb}{1.000,0.965,0.561}
\definecolor{khaki2}{rgb}{0.933,0.902,0.522}
\definecolor{khaki3}{rgb}{0.804,0.776,0.451}
\definecolor{khaki4}{rgb}{0.545,0.525,0.306}
\definecolor{LightGoldenrod1}{rgb}{1.000,0.925,0.545}
\definecolor{LightGoldenrod2}{rgb}{0.933,0.863,0.510}
\definecolor{LightGoldenrod3}{rgb}{0.804,0.745,0.439}
\definecolor{LightGoldenrod4}{rgb}{0.545,0.506,0.298}
\definecolor{LightYellow1}{rgb}{1.000,1.000,0.878}
\definecolor{LightYellow2}{rgb}{0.933,0.933,0.820}
\definecolor{LightYellow3}{rgb}{0.804,0.804,0.706}
\definecolor{LightYellow4}{rgb}{0.545,0.545,0.478}
\definecolor{yellow1}{rgb}{1.000,1.000,0.000}
\definecolor{yellow2}{rgb}{0.933,0.933,0.000}
\definecolor{yellow3}{rgb}{0.804,0.804,0.000}
\definecolor{yellow4}{rgb}{0.545,0.545,0.000}
\definecolor{gold1}{rgb}{1.000,0.843,0.000}
\definecolor{gold2}{rgb}{0.933,0.788,0.000}
\definecolor{gold3}{rgb}{0.804,0.678,0.000}
\definecolor{gold4}{rgb}{0.545,0.459,0.000}
\definecolor{goldenrod1}{rgb}{1.000,0.757,0.145}
\definecolor{goldenrod2}{rgb}{0.933,0.706,0.133}
\definecolor{goldenrod3}{rgb}{0.804,0.608,0.114}
\definecolor{goldenrod4}{rgb}{0.545,0.412,0.078}
\definecolor{DarkGoldenrod1}{rgb}{1.000,0.725,0.059}
\definecolor{DarkGoldenrod2}{rgb}{0.933,0.678,0.055}
\definecolor{DarkGoldenrod3}{rgb}{0.804,0.584,0.047}
\definecolor{DarkGoldenrod4}{rgb}{0.545,0.396,0.031}
\definecolor{RosyBrown1}{rgb}{1.000,0.757,0.757}
\definecolor{RosyBrown2}{rgb}{0.933,0.706,0.706}
\definecolor{RosyBrown3}{rgb}{0.804,0.608,0.608}
\definecolor{RosyBrown4}{rgb}{0.545,0.412,0.412}
\definecolor{IndianRed1}{rgb}{1.000,0.416,0.416}
\definecolor{IndianRed2}{rgb}{0.933,0.388,0.388}
\definecolor{IndianRed3}{rgb}{0.804,0.333,0.333}
\definecolor{IndianRed4}{rgb}{0.545,0.227,0.227}
\definecolor{sienna1}{rgb}{1.000,0.510,0.278}
\definecolor{sienna2}{rgb}{0.933,0.475,0.259}
\definecolor{sienna3}{rgb}{0.804,0.408,0.224}
\definecolor{sienna4}{rgb}{0.545,0.278,0.149}
\definecolor{burlywood1}{rgb}{1.000,0.827,0.608}
\definecolor{burlywood2}{rgb}{0.933,0.773,0.569}
\definecolor{burlywood3}{rgb}{0.804,0.667,0.490}
\definecolor{burlywood4}{rgb}{0.545,0.451,0.333}
\definecolor{wheat1}{rgb}{1.000,0.906,0.729}
\definecolor{wheat2}{rgb}{0.933,0.847,0.682}
\definecolor{wheat3}{rgb}{0.804,0.729,0.588}
\definecolor{wheat4}{rgb}{0.545,0.494,0.400}
\definecolor{tan1}{rgb}{1.000,0.647,0.310}
\definecolor{tan2}{rgb}{0.933,0.604,0.286}
\definecolor{tan3}{rgb}{0.804,0.522,0.247}
\definecolor{tan4}{rgb}{0.545,0.353,0.169}
\definecolor{chocolate1}{rgb}{1.000,0.498,0.141}
\definecolor{chocolate2}{rgb}{0.933,0.463,0.129}
\definecolor{chocolate3}{rgb}{0.804,0.400,0.114}
\definecolor{chocolate4}{rgb}{0.545,0.271,0.075}
\definecolor{firebrick1}{rgb}{1.000,0.188,0.188}
\definecolor{firebrick2}{rgb}{0.933,0.173,0.173}
\definecolor{firebrick3}{rgb}{0.804,0.149,0.149}
\definecolor{firebrick4}{rgb}{0.545,0.102,0.102}
\definecolor{brown1}{rgb}{1.000,0.251,0.251}
\definecolor{brown2}{rgb}{0.933,0.231,0.231}
\definecolor{brown3}{rgb}{0.804,0.200,0.200}
\definecolor{brown4}{rgb}{0.545,0.137,0.137}
\definecolor{salmon1}{rgb}{1.000,0.549,0.412}
\definecolor{salmon2}{rgb}{0.933,0.510,0.384}
\definecolor{salmon3}{rgb}{0.804,0.439,0.329}
\definecolor{salmon4}{rgb}{0.545,0.298,0.224}
\definecolor{LightSalmon1}{rgb}{1.000,0.627,0.478}
\definecolor{LightSalmon2}{rgb}{0.933,0.584,0.447}
\definecolor{LightSalmon3}{rgb}{0.804,0.506,0.384}
\definecolor{LightSalmon4}{rgb}{0.545,0.341,0.259}
\definecolor{orange1}{rgb}{1.000,0.647,0.000}
\definecolor{orange2}{rgb}{0.933,0.604,0.000}
\definecolor{orange3}{rgb}{0.804,0.522,0.000}
\definecolor{orange4}{rgb}{0.545,0.353,0.000}
\definecolor{DarkOrange1}{rgb}{1.000,0.498,0.000}
\definecolor{DarkOrange2}{rgb}{0.933,0.463,0.000}
\definecolor{DarkOrange3}{rgb}{0.804,0.400,0.000}
\definecolor{DarkOrange4}{rgb}{0.545,0.271,0.000}
\definecolor{coral1}{rgb}{1.000,0.447,0.337}
\definecolor{coral2}{rgb}{0.933,0.416,0.314}
\definecolor{coral3}{rgb}{0.804,0.357,0.271}
\definecolor{coral4}{rgb}{0.545,0.243,0.184}
\definecolor{tomato1}{rgb}{1.000,0.388,0.278}
\definecolor{tomato2}{rgb}{0.933,0.361,0.259}
\definecolor{tomato3}{rgb}{0.804,0.310,0.224}
\definecolor{tomato4}{rgb}{0.545,0.212,0.149}
\definecolor{OrangeRed1}{rgb}{1.000,0.271,0.000}
\definecolor{OrangeRed2}{rgb}{0.933,0.251,0.000}
\definecolor{OrangeRed3}{rgb}{0.804,0.216,0.000}
\definecolor{OrangeRed4}{rgb}{0.545,0.145,0.000}
\definecolor{red1}{rgb}{1.000,0.000,0.000}
\definecolor{red2}{rgb}{0.933,0.000,0.000}
\definecolor{red3}{rgb}{0.804,0.000,0.000}
\definecolor{red4}{rgb}{0.545,0.000,0.000}
\definecolor{DeepPink1}{rgb}{1.000,0.078,0.576}
\definecolor{DeepPink2}{rgb}{0.933,0.071,0.537}
\definecolor{DeepPink3}{rgb}{0.804,0.063,0.463}
\definecolor{DeepPink4}{rgb}{0.545,0.039,0.314}
\definecolor{HotPink1}{rgb}{1.000,0.431,0.706}
\definecolor{HotPink2}{rgb}{0.933,0.416,0.655}
\definecolor{HotPink3}{rgb}{0.804,0.376,0.565}
\definecolor{HotPink4}{rgb}{0.545,0.227,0.384}
\definecolor{pink1}{rgb}{1.000,0.710,0.773}
\definecolor{pink2}{rgb}{0.933,0.663,0.722}
\definecolor{pink3}{rgb}{0.804,0.569,0.620}
\definecolor{pink4}{rgb}{0.545,0.388,0.424}
\definecolor{LightPink1}{rgb}{1.000,0.682,0.725}
\definecolor{LightPink2}{rgb}{0.933,0.635,0.678}
\definecolor{LightPink3}{rgb}{0.804,0.549,0.584}
\definecolor{LightPink4}{rgb}{0.545,0.373,0.396}
\definecolor{PaleVioletRed1}{rgb}{1.000,0.510,0.671}
\definecolor{PaleVioletRed2}{rgb}{0.933,0.475,0.624}
\definecolor{PaleVioletRed3}{rgb}{0.804,0.408,0.537}
\definecolor{PaleVioletRed4}{rgb}{0.545,0.278,0.365}
\definecolor{maroon1}{rgb}{1.000,0.204,0.702}
\definecolor{maroon2}{rgb}{0.933,0.188,0.655}
\definecolor{maroon3}{rgb}{0.804,0.161,0.565}
\definecolor{maroon4}{rgb}{0.545,0.110,0.384}
\definecolor{VioletRed1}{rgb}{1.000,0.243,0.588}
\definecolor{VioletRed2}{rgb}{0.933,0.227,0.549}
\definecolor{VioletRed3}{rgb}{0.804,0.196,0.471}
\definecolor{VioletRed4}{rgb}{0.545,0.133,0.322}
\definecolor{magenta1}{rgb}{1.000,0.000,1.000}
\definecolor{magenta2}{rgb}{0.933,0.000,0.933}
\definecolor{magenta3}{rgb}{0.804,0.000,0.804}
\definecolor{magenta4}{rgb}{0.545,0.000,0.545}
\definecolor{orchid1}{rgb}{1.000,0.514,0.980}
\definecolor{orchid2}{rgb}{0.933,0.478,0.914}
\definecolor{orchid3}{rgb}{0.804,0.412,0.788}
\definecolor{orchid4}{rgb}{0.545,0.278,0.537}
\definecolor{plum1}{rgb}{1.000,0.733,1.000}
\definecolor{plum2}{rgb}{0.933,0.682,0.933}
\definecolor{plum3}{rgb}{0.804,0.588,0.804}
\definecolor{plum4}{rgb}{0.545,0.400,0.545}
\definecolor{MediumOrchid1}{rgb}{0.878,0.400,1.000}
\definecolor{MediumOrchid2}{rgb}{0.820,0.373,0.933}
\definecolor{MediumOrchid3}{rgb}{0.706,0.322,0.804}
\definecolor{MediumOrchid4}{rgb}{0.478,0.216,0.545}
\definecolor{DarkOrchid1}{rgb}{0.749,0.243,1.000}
\definecolor{DarkOrchid2}{rgb}{0.698,0.227,0.933}
\definecolor{DarkOrchid3}{rgb}{0.604,0.196,0.804}
\definecolor{DarkOrchid4}{rgb}{0.408,0.133,0.545}
\definecolor{purple1}{rgb}{0.608,0.188,1.000}
\definecolor{purple2}{rgb}{0.569,0.173,0.933}
\definecolor{purple3}{rgb}{0.490,0.149,0.804}
\definecolor{purple4}{rgb}{0.333,0.102,0.545}
\definecolor{MediumPurple1}{rgb}{0.671,0.510,1.000}
\definecolor{MediumPurple2}{rgb}{0.624,0.475,0.933}
\definecolor{MediumPurple3}{rgb}{0.537,0.408,0.804}
\definecolor{MediumPurple4}{rgb}{0.365,0.278,0.545}
\definecolor{thistle1}{rgb}{1.000,0.882,1.000}
\definecolor{thistle2}{rgb}{0.933,0.824,0.933}
\definecolor{thistle3}{rgb}{0.804,0.710,0.804}
\definecolor{thistle4}{rgb}{0.545,0.482,0.545}
\definecolor{gray5}{rgb}{0.051,0.051,0.051}
\definecolor{gray10}{rgb}{0.102,0.102,0.102}
\definecolor{gray15}{rgb}{0.149,0.149,0.149}
\definecolor{gray20}{rgb}{0.200,0.200,0.200}
\definecolor{gray25}{rgb}{0.251,0.251,0.251}
\definecolor{gray30}{rgb}{0.302,0.302,0.302}
\definecolor{gray35}{rgb}{0.349,0.349,0.349}
\definecolor{gray40}{rgb}{0.400,0.400,0.400}
\definecolor{gray45}{rgb}{0.451,0.451,0.451}
\definecolor{gray50}{rgb}{0.498,0.498,0.498}
\definecolor{gray55}{rgb}{0.549,0.549,0.549}
\definecolor{gray60}{rgb}{0.600,0.600,0.600}
\definecolor{gray65}{rgb}{0.651,0.651,0.651}
\definecolor{gray70}{rgb}{0.702,0.702,0.702}
\definecolor{gray75}{rgb}{0.749,0.749,0.749}
\definecolor{gray80}{rgb}{0.800,0.800,0.800}
\definecolor{gray85}{rgb}{0.851,0.851,0.851}
\definecolor{gray90}{rgb}{0.898,0.898,0.898}
\definecolor{gray95}{rgb}{0.949,0.949,0.949}
\definecolor{gray100}{rgb}{1.000,1.000,1.000}
\definecolor{DarkGray}{rgb}{0.663,0.663,0.663}
\definecolor{DarkBlue}{rgb}{0.000,0.000,0.545}
\definecolor{DarkCyan}{rgb}{0.000,0.545,0.545}
\definecolor{DarkMagenta}{rgb}{0.545,0.000,0.545}
\definecolor{DarkRed}{rgb}{0.545,0.000,0.000}
\definecolor{LightGreen}{rgb}{0.565,0.933,0.565}

\psset{unit=1mm}

\begin{document}

\title{\vspace{-3cm}{\bf {\color{Maroon}{Resource allocation  with costly participation}}\thanks{The first draft October 2010. This version December 2015.}
}}

\author{{\bf Ali Kakhbod}\thanks{Department of Economics, Massachusetts Institute of Technology (MIT), Cambridge, MA 02139, USA. Email: \url{akakhbod@mit.edu}.}
\\
{Massachusetts Institute of Technology}
      }

\date{\small{}}

\maketitle
\thispagestyle{empty}

%
%
%
\begin{abstract}
We propose a new all-pay auction format in which risk-loving  bidders pay a constant fee each time they bid for an object whose monetary value is common knowledge among the bidders, and  bidding fees are  the only source of benefit for the seller. We show that for the proposed model there exists a {unique} Symmetric Subgame Perfect Equilibrium (SSPE). The characterized SSPE is stationary when re-entry in the auction is allowed, and it is Markov perfect when re-entry is forbidden. Furthermore, we fully characterize the expected revenue of the seller.  Generally, with or without re-entry, it is more beneficial for the seller to choose $v$ (value of the object), $s$ (sale price), and $c$ (bidding fee) such that $\frac{v-s}{c}$ becomes sufficiently large. In particular, when re-entry is permitted:  the expected revenue of the seller is \emph{independent} of the number of bidders,  decreasing in the sale price, increasing in the value of the object, and  decreasing in the bidding fee; Moreover,   the seller's revenue is equal to the value of the object when players are risk neutral, and it is strictly greater than the value of the object when bidders are risk-loving.   We further show that allowing re-entry can be important in practice. Because, if the seller were to run such an auction without allowing re-entry, the auction would last a long time, and for almost all of its duration have only two remaining players. Thus, the seller's revenue relies on those two players being willing to participate, without any breaks, in an auction that might last for thousands of rounds.\\
%
%
%
%

\noindent\textbf{Keywords:}
Pay-to-bid auctions, risk-loving bidders, risk neutral bidders, revenue management.\\

\noindent\textbf{JEL Classification:}
 C61, C73, D44

\end{abstract}

\newpage
\begin{section}{Introduction}
Over the past few years, certain internet auctions, commonly referred to as pay-to-bid auctions, have seen a rapid rise in popularity.  The basic structure of the auction
is as follows: (i) The price starts at zero. (ii) Each bid costs 1 cent for any  bidder.
(iii) Each bid adds 10 seconds to the game clock, so  the auction never ends while there are still willing bidders. (iv) If the clock reaches zero, the final bidder wins the object and pays the sell price.
Due to the structure of these auctions, when the price is low, each player prefers to bid if afterwards no opponent will enter before the clock runs out. However, if players are too likely to enter in the future, no player will want to bid.

The mechanic of pay-to-bid auctions (also known as penny auctions) has parallels with the all-pay auctions and the war-of-attrition game format, \cite{fudenberg}.  Because  with the classic all-pay auction,  \cite{shubik}, all bidders pay a positive sum if they choose to participate in the auction, regardless of whether they win or lose. The costly action, required to become the winning bidder, is also reminiscent of the war-of-attrition game format. 
 {All-pay auction has been applied to describe rent-seeking, political contests, R\&D races, and job-promotions. Under full information in one-shot (first price) all-pay auctions, the full characterization of equilibria is given by \cite{Baye}. A second-price all-pay auction, also called war of attrition, was proposed  by \cite{Smith} in the context of theoretical biology and, under full information,  the full characterization of equilibria of it is given by \cite{Hen1}. The war of attrition game has been applied to many problems in economics, most notably industrial competition (e.g. \cite{fudenberg2}), public goods provision (e.g. \cite{Bliss}) and Bargaining (e.g. \cite{ordover} and \cite{rub2}). Recently,  \cite{Siegel} provides an equilibrium payoff characterization for general class of all-pay contests.}

In  pay-to-bid auctions, since the main cost incurred by bidders come in the form of bidding fees (which are individually small),  bidders are not required to place a bid every round in order to stay in the auction   (that is, entry  and re-entry at any round of the auction is allowed). In contrast,  re-entry in all-pay auctions is forbidden. Therefore, all-pay auctions never allow the actual winner to pay less than the losers, but in pay-to-bid auctions it happens, in practice, relatively often\footnote{In more detail, in both  all-pay auctions (in particular war of attrition) and pay-to-bid auctions, players must pay a cost for the game to continue and a player wins when other players decide not to pay this cost. However, they are different in one notable way that in a pay-to-bid auction each player can bid in any period  independent of the history of the game, unlike in a war of attrition, in which all players who continue to play must pay a cost in each period. Therefore,  unlike in pay-to-bid auctions,  in war of attrition  actual winner never pay less than the losers.}.

What really sets this particular auction format apart from other nontraditional auctions is the success of its real world implementation which appears to be highly profitable for the website operator/auctioneer (who is the sole seller of goods).
{In December 2008, 14 new websites conducted such auctions; by November 2009, the format had increased to 35 websites. Over the same time span, traffic among these sites has increased from 1.2 million to 3.0 million unique visitors per month. 
 For comparison, traffic at ebay.com fluctuated around 75 million unique visitors per month throughout that period.  Pay-to-bid auctions have garnered 4\% of the traffic held by the undisputed leader in
online auctions, \cite{platt}.}

\begin{subsection}{Risk-loving behavior, Having common valuation, and \\
Swoopo's  bankruptcy}
These internet auctions are  a form of gambling. The bidder
deposits a small fee to play, aspiring to a big payoff  of obtaining the item well below its value, with a major difference that the probabilities of winning are endogenous. Due to the gambling feature of these auctions and the fact that these internet auctions  prominently advertise themselves  as ``entertainment shopping'',  the risk-loving behavior (i.e.  having a preference for  risk)   by participants of these auctions is natural\footnote{For example, for video game systems (such as the  Playstation, Wii, or Xbox) the bidders may not be able to justify to their spouse, their parent, or themselves  spending \$450 on a Xbox at a retail store; however, the potential to win the Xbox early in the auction for only a fraction of that makes it worth the 1 cent  gamble, even at unfair odds.}. 

Generally, in these internet auctions  the valuation of the item is known and the same among all potential bidders.  Since all items are new, unopened, and readily available from  internet retailers, the market prices of these items are well established.
In fact, one could imagine that   the value of the object  is the lowest price for which the item may be obtained elsewhere. This feature, unlike in the first and second-price auctions, is quite common in all-pay or war-of-attrition auctions, which are the closest relatives of the pay-to-bid auction.  

Swoopo.com is one of such websites that was initially very successful.  According to an August 2009 article from \emph{The Economist}, Swoopo has 2.5 million registered users and earned 32 million dollars in revenue, in 2008. Data collected by a blogger shows that over April and May of 2009 Swoopo sold items for an average of 188\% of their listed value. On 26 March 2011, Swoopo's parent company filed for bankruptcy and shut down the auction
website. On 8 February 2012, DealDash the longest running pay-to-bid auction website in the U.S. acquired the domain Swoopo.com and the URL currently redirects to DealDash's own website. The exit of Swoopo might suggest: (i) demand for pay-to-bid auctions fell drastically. (ii) Swoopo was mismanaged and made a set of poor strategic choices.   These effects are empirically disentangled in \cite{ned}.   In early 2008, when penny auctions are launched, there are very few online pay-to-bid auctions, and  essentially Swoopo acts as a monopoly.  Over time, by increasing the number of visitors, more entrants come into the market, reducing  the level of concentration. When Swoopo exits, the number of market monthly visitors drops \emph{temporarily}, but  soon rises back to around 0.0075 of Internet traffic, \cite{ned}. This result, therefore,  suggests that after Swoopo's exit demand for pay-to-bid auctions has continued to remain high.


\end{subsection}

\begin{subsection}{Contribution of the paper}\label{cont}
This work, in continuation of \cite{kakhbod}, focuses  on risk-loving players with Constant Absolute Risk Loving (CARL) utilities. We  present  a stylized model of a pay-to-bid auction  with the following properties:  (i)  Biding fee is small with respect to the value of the object. (ii)  Re-entry in the auction may be allowed (i.e.,  all the original players can participate in any round of the game) or forbidden (ie,  in each round of the game only players  who have participated in the previous round can participate).

 We first establish that, with or without re-entry permission,  there exists a unique  symmetric subgame perfect equilibrium, which is \emph{stationary} when re-entry is permitted, and it is  \emph{Markov perfect}\footnote{A Markov perfect equilibrium is a profile of Markov strategies that yields a Nash equilibrium in every proper subgame. A Markov strategy is one that does not depend at all on variables that are functions of the history of the game except those that affect payoffs.} when re-entry is forbidden. 

As shown in \cite{kakhbod}, when re-entry is permitted,  a closed form expression of the seller's revenue is computable that becomes {independent} of the number of players. The closed form expression reveals that the  seller's revenue is increasing  in the value of the object,   decreasing in the biding fee, decreasing the sale price, and  decreasing in the risk-loving coefficient. These results are compatible with the relatively high valued objects sold in the online pay-to-bid auctions with a low bid fee and a low sale price.  Moreover, the seller's revenue is exactly equal to the value of the object, when players are risk neutral, and is strictly greater than the value of the object, when players are risk-loving.  Furthermore, for sufficiently small biding fee, it  is strictly \emph{greater} than a seller's revenue from a standard lottery, for any number of players.
 
Here, we show, generally, with or without re-entry, it is more beneficial for the seller to choose $v$ (value of the object), $s$ (sale price), and $c$ (bidding fee) so that $\frac{v-s}{c}$ becomes sufficiently large.  Moreover, whether or not entry or re-entry is permitted  in any round of the game,  the strategy that each agent follows in equilibrium  is independent of her wealth/budget level. Furthermore,   in any sub-game starting from round $t, t\in \{1,2,3,\cdots\}$, the expected utility of each  player $i$, whose budget level in round $t$ is $w_{i,t}$,  is exactly equal to $u(w_{i,t})$, where $u(\cdot)$ denotes the player's utility.

We show that allowing re-entry might be important in practice  compared to the scenario in which re-entry is forbidden. Recall that the seller's expected revenue is increasing with respect to the value of the object and decreasing with respect to the bid fee and the sale price. Given this result, we establish that, in the limit, as the value of the object grows with respect to the bid fee and the sale price, the expected length of the game tends to infinity.  Furthermore, with probability approaching to one, in the game in which re-entry is forbidden, the number of players will be reduced to only two remaining players that play the game until it ends. Moreover, the expected time in which the number of players reduces to two is \emph{negligible} in comparison to the entire expected length of the game.

In sum, when a high valued object is sold with a low bid fee and 
a low sale price,  an auction in which re-entry is forbidden not only will last on average for a very long time, but also most likely it will  only have two players who play for almost the entire duration of the game. Therefore, because the auction will end, if either of the two remaining players chooses to opt out, due to   an exogenous reason,  it is easy to see why the auction in which re-entry is forbidden  is \emph{undesirable} from the seller's point view. 
\end{subsection}

\begin{subsection}{Organization of the paper }
The rest of the paper is organized as follows. In Section \ref{sec1} the model is precisely described. In Section \ref{compare} we discuss about related literature. In Section \ref{prop} we present the properties of the proposed model. We conclude in Section \ref{con}.\\
\end{subsection}
\end{section}

\begin{section}{The Model}
\label{sec1}
Consider the following dynamic  game with complete information involving an object with monetary value $v$ and $n$ strategic bidder/agent/player/participant.
 In each round $t, t\in \{1,2,3,\cdots\}$, of the auction/game player $i$  chooses an action from the set of pure strategies/messages $S=\{\mbox{Bid, No Bid}\},$ $s_{i,t} \in S,$ and  observes her opponents actions, $s_{j,t}, j\neq i$. In each round of the game, players submit their messages simultaneously. 
 Playing $\{\mbox{Bid}\}$ is costly. Each player immediately pays $c<v$ dollars (the bid fee) to the seller each time she plays $\{\mbox{Bid}\}$.  In any particular round, If a player is the sole player who plays $\{\mbox{Bid}\}$, she wins the object and pays the sale price. The sale price is fixed and denoted by $s$. In any round of the game, if more than one player play $\{\mbox{Bid}\}$, the game continues in the next round. 
 
  It may arise a difficulty that no participant plays $\{\mbox{Bid}\}$, in a particular round of the game, that  we resolve it by the following assumption.
 
\begin{Assumption}[Tie Breaking]
\label{fud}
We assume if in a round of the game  \emph{all} the  players play $\{\mbox{No Bid}\}$, they resubmit their messages, ie,  they re-play that period of the game.
\end{Assumption}

For this model, we consider the following two scenarios: 
\begin{itemize}
\item[(i)] The game with re-entry option. In this scenario, all the original players are allowed to participate/bid at any round of the game, regardless of the history of the game and the way that they behaved. 
\item[(ii)] The game without re-entry option. In this scenario, the set of participating players in the next round of the game are the ones who have bid. That is, in any round of the game,  a player, who does not bid, will be out forever.
\end{itemize}

\end{section}
\medskip
\begin{section}{Related work}
\label{compare}

There has been a great deal of recent interest in pay-to-bid auctions. Most of this work offers the large revenues earned by websites like Dealdash.com (see \cite{byers,ned,hinosar,platt,kakhbod}). 
The  proposed formulation in our paper  is different from the formulations in \cite{hinosar,ned,platt}  in one notable way.  In \cite{hinosar,ned,platt}, the authors assume that, in each round $t$ of the game, a \emph{leader} is selected randomly  from the players participating in round $t$ (ie, the leader is chosen randomly from the players who chose to play $\{\mbox{Bid}\}$ in round $t$). And the leader wins the object,  if non of the remaining players  play $\{\mbox{Bid}\}$ in the next round of the game.  This formulation differs from ours.  Because  a player, in our formulation, wins the object,  if no other player bids (ie, no other player play $\{\mbox{Bid}\}$ in \emph{that particular round})\footnote{We note that the formulation in \cite{hinosar,ned,platt} models the online websites, like Dealdash.com, more accurately, but as \cite{hinosar} establishes, the  equilibrium analysis is \emph{completely intractable}. Therefore, in \cite{hinosar,ned,platt} the authors focus only on  Markov equilibria,  where players instead of conditioning their behavior on the whole histories of the game, they only condition their behavior on the current state of the game, whereas in our formulation  we do not have such assumption. We also note that, the models in \cite{hinosar,ned,platt} are also different from one another. For example, in \cite{hinosar}  and \cite{platt}  (similar to ours), whenever a player plays $\{\mbox{Bid}\}  $ she pays the bidding fee, where as \cite{ned}  assumes that only the submitting bidder who was chosen to be the next leader has to incur the bid cost}.


 Another notable difference of our model and the models in \cite{byers,hinosar,ned}  is that, similar to \cite{platt},   we focus on risk-loving players while in \cite{byers,hinosar,ned} the authors assume players are risk neutral. In this respect, our paper is closely related to \cite{platt}.  But it is, also, different from \cite{platt}, in formulation of the model, such that 
 we  get a more tractable model with  a \emph{unique} symmetric subgame perfect equilibrium, where in \cite{platt},  the focus is on  stationary equilibria, and  there are multiple symmetric equilibria. Another difference of our work and \cite{platt} (and \cite{byers,hinosar,ned}) is that the main focus of our paper is to study the effect of having re-entry in any round of the auction, in comparison to the case where re-entry is forbidden.


We further note that the models in \cite{hinosar,ned,platt} are also different from one another. For example, in \cite{hinosar}  and \cite{platt}  (similar to ours) whenever a player plays $\{\mbox{Bid}\}  $ she pays the bidding fee, however,  \cite{ned}  assumes that only the submitting bidder, who was chosen to be the next leader, has to incur the bid cost.

In pay-to-bid auctions  each player is aware of the number of current players at each round of the game.  We will also assume that the number of players is common knowledge. In \cite{byers}, the authors investigate a situation in which players are not aware of the exact number of  players, and they improperly estimate the number of participants. This assumption proves to have dramatic impact on the analysis and, in particular, on the expected revenue of the seller.

It is often assumed that incurred or sunk costs do not directly affect individual's decision but may have an effect on future decisions\footnote{There is a large body of  literature in psychology and behavioral economics that address this issue.  For example, previous expenditures can affect future decisions by changing one's remaining disposable income.  The sunk cost fallacy is naturally connected to pay-to-bid auctions. This is due to the fact that  in these auctions players/participants spend more and more money as time goes on, and participants who suffer from the sunk cost fallacy will feel a greater and greater need to justify their loses by winning the prize.}. In \cite{ned}, the author has built a model of pay-to-bid auction,  similar to \cite{hinosar},  in which bidders suffer from the {sunk cost fallacy}. Further, bidders behave so na\"\i vely in the sense that they do not foresee the effect of their loses on their future preferences.  Therefore,  bidders overbid more as they become more monetarily invested in the auction.   
\end{section}



%
%

\begin{section}{Properties of the Game}\label{prop}
In this section, we investigate the properties of the model proposed in section \ref{sec1}. We first define the utility function of the players.

\begin{subsection}{Constant Absolute Risk-Loving Utility Function}\label{utility}
A Von Neumann-Morganstern utility function $u:\mathbb{R}\rightarrow\mathbb{R}$ is said to be  Constant Absolute Risk Loving (CARL), if the Arrow-Pratt measure of risk 
$$
R(x)=-\frac{u''(x)}{u'(x)}
$$
is equal to some constant $\rho$ ($\rho<0$, called risk-loving coefficient) for all $x$. Thus, any CARL utility function has the \emph{unique} following form (up to an affine transformation)
\begin{eqnarray}
\label{ut}
u(x)=\frac{1-e^{-\rho x}}{\rho}+K.
\end{eqnarray}
 \\
We consider players are risk-loving and, therefore, their utility functions are in the form of \eqref{ut}. 
For simplicity, we focus on the fundamental form in \eqref{ut} so that $u(0)=0$, ie, $K=0$.  Further, we assume  players do not discount future consumption.

\begin{Remark}
In the limit when $\rho$ tends to 0, the utility function in \eqref{ut} converges (point wise) to the risk neutral utility function $u(x)=x$.  
\end{Remark}
\end{subsection}

%
\begin{subsection}{Equilibrium analysis}

In the following theorem, we establish that for the model proposed in Section \ref{sec1} there is a \emph{unique} Symmetric\footnote{In this paper we focus on characterizing symmetric equilibria.  There might be asymmetric equilibria but is not of our interest. In Corollary \ref{cor1} we comment on the appropriateness of analyzing symmetric equilibria.} Sub-game Perfect Equilibrium (SSPE), which is \emph{stationary}\footnote{It is stationary since the strategy each player follows in each round $t$ of the game  depends only on the relevant state variables: the number of players $n$, the sale price $s$, the bid fee $c$, and the object's value $v$.}  when re-entry is allowed, and it is  Markov perfect\footnote{It is Markov perfect since the strategy each player follows in each round $t$ of the game  depends only on the relevant state variables:  number of  remaining players $n_t$, the sale price $s$, the bid fee $c$, and the object's value $v$. We further note that, a Markov perfect equilibrium is a profile of Markov strategies that yields a Nash equilibrium in every proper subgame. A Markov strategy is one that does not depend at all on variables that are functions of the history of the game except those that affect payoffs.} when re-entry is forbidden.

\begin{Theorem}
\label{th1}
 In any \emph{symmetric} sub-game perfect equilibrium of the game proposed in section \ref{sec1}, with $n\geq 2$ players,  the following properties are satisfied.
\begin{itemize}
\item[I.] Both with and without re-entry, in any sub-game starting from round $t, t\in \{1,2,3,\cdots\}$, the expected utility of  player $i$, whose wealth/budget level  is $w_{i,t}$,  is exactly equal to $u(w_{i,t})$.
\item[II.] When re-entry is allowed,  in each period, each player \emph{purely} randomizes over $\{\mbox{Bid, No Bid}\}$ and chooses to play $\{\mbox{Bid}\}$ with the following \emph{stationary} probability:
\begin{eqnarray*}
\label{see1}
1- \sqrt[n-1]{\frac{u(c)}{u(v-s)}}.
\end{eqnarray*}	
Moreover,  the  symmetric sub-game perfect equilibrium is  \emph{unique} and \emph{stationary}.
\item[III.] When re-entry is \emph{not} allowed,  in each period, each player \emph{purely} randomizes over $\{\mbox{Bid, No Bid}\}$ and chooses to play $\{\mbox{Bid}\}$ with the following  probability:
\begin{eqnarray*}
\label{see1}
1- \sqrt[n_t-1]{\frac{u(c)}{u(v-s)}},
\end{eqnarray*}	
where  $n_t$ is the number of players in round $t$. 

Moreover,  the  symmetric sub-game perfect equilibrium is  \emph{unique} and \emph{Markov perfect}.
\end{itemize}
\end{Theorem}
\begin{proof} See Appendix.
\end{proof}
From the above theorem, we directly obtain the following  corollaries.

\begin{Corollary}\label{cor1} Since the decision taken by the players is \emph{independent} of their wealth level, we are able to focus on the \emph{symmetric} sub-game perfect equilibria despite wealth asymmetries.   
 \end{Corollary}
 
  \begin{Corollary} Both with and without re-entry,  in each period of the game,  the probability of playing  $\{\mbox{Bid}\}$  is {strictly} greater  than zero and {strictly} less than one. That is,  in each period of the game, each player is \emph{indifferent} between playing $\{\mbox{Bid}\}$ and $\{\mbox{No Bid}\}$.
 \end{Corollary}
 
 \begin{Corollary} 
Theorem \ref{th1} is also hold, when players are risk-neutral\footnote{The risk neutral case can be easily implied by taking a limit  $\rho \rightarrow 0$ in \eqref{ut}, that implies (point-wise) $u(x)=x$ for any $x$.}.  In particular, when players are risk neutral,  in any round $t, t\in \{1,2,3,\cdots\},$ of the game:  \emph{(1)}  When re-entry is allowed, each player  chooses to play $\{\mbox{Bid}\}$ with the \emph{stationary} probability $1- \sqrt[n-1]{\frac{c}{v-s}}$.  \emph{(2)}  When re-entry is \emph{not} allowed, each player  chooses to play $\{\mbox{Bid}\}$ with the probability $1- \sqrt[n_t-1]{\frac{c}{v-s}}$.
  \end{Corollary}
  
    \begin{Corollary} Both with and without re-entry, the probability that any bidding player\footnote{Bidding player means a player who plays $\{\mbox{Bid}\}$.} wins the object, in any round of the game, is  equal to $\frac{u(c)}{u(v-s)}$.
 \end{Corollary}

 \end{subsection}
 
 \begin{subsection}{Revenue analysis}\label{revenue00}

In Theorem \ref{th3}, we focus on computing the seller's profit, in equilibrium, only from the bid fees.

\begin{Theorem}
\label{th3}
Both with and without re-entry, in the unique SSPE characterized in Theorem \ref{th1}, the expected earning  of the seller only from the bid fees is exactly equal to
\begin{eqnarray}
\label{general}
\sum_{t=1}^{\infty}\left[\frac{n_tp_t}{1-(1-p_t)^{n_t}} \times c \times \prod_{s=1}^{t-1}\left(1-\frac{n_sp_s(1-p_s)^{n_s-1}}{1-(1-p_s)^{n_s}}\right)\right],
\end{eqnarray}
where $n_t$ denotes the number of remaining players. Note that when re-entry is permitted $n_t=n$, for all $t$.
\end{Theorem}
\begin{proof} See Appendix.
\end{proof}

\begin{Theorem}\label{th333}
For a given bidding fee $c>0$,  both with and without re-entry, in the unique SSPE characterized in Theorem \ref{th1},  when $\frac{v-s}{c}$ tends to infinity, the expected profit of the seller is maximized.  
\end{Theorem}

\begin{proof} See Appendix.
\end{proof}

This result is intuitive because risk-loving bidders prefer low probability and potentially very lucrative gambles  to gamble that are equal in expectations but have lower variance. Moreover, the seller is able to earn large profits even when the bid fee and the sale price are small compared to the object's value  because the expected number of bids and the expected length of these auctions are large enough to compensate this discrepancy. We investigate more about  this property in Theorem \ref{th4}.

When re-entry is allowed,
since the SSPE described in Theorem \ref{th1} becomes \emph{stationary},  it is doable to simplify Eq. \eqref{general} and derive its closed form. As a result, we will be able to derive a few interesting insights from its expression.  As shown in \cite{kakhbod}, these insights are summarized as follows. The expected revenue of the seller  is decreasing in the sale price $(s)$, increasing in the value of the object $(v)$, and decreasing in the bidding fee $(c)$. Moreover,  the expected revenue of the seller is independent of the number of players.  In addition,  for any number of players, with sufficiently small biding fee,  the seller' revenue  is strictly \emph{greater} than seller's revenue from any standard lottery\footnote{The standard lottery is defined as follows. A  seller offers lottery tickets for an object of (monetary) value $v$ to $n$ potential buyers with CARL utility. Cost of buying a ticket for each player is equal to $c>0$, and the winner is chosen randomly.}.

To be precise, we have the following theorem.

\begin{Theorem}[\cite{kakhbod}]
\label{th2}
Consider the SSPE of the {Bid-No Bid} game  when  re-entry is permitted. \\
If players are risk neutral (ie, $\rho=0$ in \eqref{ut}),  then 
\begin{itemize}
\item[I.] The expected revenue of the seller is  equal to the value of the object  $(v)$.
\end{itemize}
If players are risk-loving (ie, $\rho<0$ in \eqref{ut}), then
\begin{itemize}
\item[II.]   The expected revenue of the seller is equal to $\frac{u(v-s)c}{u(c)}+s$.
\item[III.]  The expected revenue of the seller is independent of the number of players.
\item[IV.]    The expected revenue of the seller is strictly greater than the value of the object  $(v)$.
\item[V.] The expected revenue of the seller is strictly increasing in the value of the object  $(v)$,  decreasing in the sale price  $(s)$,   decreasing in the biding fee $(c)$, and  decreasing in $\rho$ (ie, the seller earns more  when players are more risk-loving).
\item[VI.] The maximum revenue the seller can earn is $u(v)=\frac{1-e^{-\rho v}}{\rho}$.
\item[V.II] For any number of players, with sufficiently small biding fee,  the seller's revenue is strictly \emph{greater} than a standard lottery. 
\end{itemize}
\end{Theorem}

\end{subsection}

%
%
%

 \begin{subsection}{With entry vs. Without re-entry}
 In this section, we investigate the effect of permitting re-entry in the game proposed in Section \ref{sec1}. 
  As shown in Section \ref{revenue00}, generally, with or without re-entry, it is more beneficial for the seller to choose $v, s,$ and $c$ such that $\frac{v-s}{c}$ becomes sufficiently large. In the following theorem, we characterize the impact of $\frac{v-s}{c}\rightarrow \infty$ on the auction when re-entry is forbidden.

 \begin{Theorem}\label{th4}
 Without re-entry option, as $\frac{v-s}{c}$ tends to infinity the following properties are hold.
\begin{itemize}
 \item[I.] The expected number of rounds of the game  tends to infinity. 
\item[II.]  When $n>2$, with probability approaching to one, the auction is reduced to one with only two remaining players before the auction ends. 
\item[III.] The expected time until $n-2$ players exit in the auction tends to zero, with respet to the   overall expected length of the auction. 
\end{itemize}
 \end{Theorem}
\begin{proof} See Appendix.
\end{proof}

  As the above theorem shows, the  expected length of the auction, without re-entry option, tends to infinity when $\frac{v-s}{c}\rightarrow \infty$. Furthermore, in an auction without re-entry, with probability approaching to one, the auction is reduced to one with only two remaining  players before the auction ends, and the expected time for this to happen becomes arbitrary small compared to the expected length of the auction. Thus, the above theorem predicts that when $\frac{v-s}{c}$ is large not only the game  last for a very long time  but also the game is likely to have only two remaining players participating in it for almost its entire duration.  Therefore, running auctions with re-entry option will be helpful to alleviate this issue because, despite in auctions without re-entry option, these auctions can be successfully implemented without any two particular players being willing to participate for the entirety of the auction.

 \end{subsection}

\end{section}


\begin{section}{Conclusion}\label{con}
%
%

We developed a dynamic pay-to-bid auction for selling an object whose monetary value is common knowledge among risk-loving bidders, and bidding fees are the only source of benefit for the seller.  We established that for the proposed model there exists a \emph{unique} symmetric subgame perfect equilibrium. The characterized equilibrium is stationary when re-entry in the auction is allowed, and it is Markov perfect when re-entry is forbidden.
Furthermore, in equilibrium,  the strategy chosen by the players is independent of their wealth level whether or not re-entry is permitted in the auction.

 When re-entry is permitted,   the expected revenue of the seller is independent of the number of buyers,  increasing in the value of the object,  decreasing in the bid fee, and decreasing in the  sale price. Furthermore, the seller's expected revenue, when bidders are risk neutral,  is  equal to the value of the object, and it is {strictly} greater than  the value of the object  when bidders are risk loving. Moreover,  the seller's revenue is strictly increasing in the degree of risk-loving.  Thus, when bidders are risk-loving, it is more {profitable} to sell expensive objects with a low bid fee and a low sale price. We further compared the seller's revenue  with a standard lottery. As a result, for sufficiently small biding fee, the seller's revenue   is {strictly higher} than a standard lottery, for any number of players.

We further showed that allowing re-entry maybe {surprisingly} important in practice. Because, if the seller were to run such an auction without allowing re-entry, the auction would last a long time, and for almost all of its duration have only two remaining players. Therefore, the seller's revenue relies on those two players being willing to participate, without any breaks, in an auction that might last for thousands of rounds.

\end{section}



\begin{thebibliography}{1}
\bibitem{fudenberg2}
D. Fudenberg, J. Tirole, \textit{A theory of exit in duopoly}. Econometrica, vol. 54pp. 934-960, 1986.
\bibitem{ordover}
J. Ordover and A. Rubinstein. \textit{A sequential concession game with asymmetric information}.
Quarterly Journal of Economics. vol. 101, pp. 879-888, 1986.
\bibitem{rub2}
L. Kornhauser, A. Rubinstein and C. Wilson \textit{Reputation and patience in the war of attrition}.
Econometrica. vol. 56, pp. 15-24, 1989.
\bibitem{Siegel}
R. Siegel. \textit{All-Pay Contests}. Econometrica, vol. 77, no. 1, pp. 71-92, 2009.
\bibitem{Hen1}
K. Hendricks, A. Weiss, and C. Wilson. \textit{The War of Attrition
in Continuous Time with Complete Information}. International Economic
Review. vol. 29, no. 4, pp. 663-680, 1988.
\bibitem{fudenberg}
D. Fudenberg, J. Tirole, \textit{Game Theory}, MIT Press, 1991.
\bibitem{Bliss}
C. Bliss and B. Rebuff. \textit{Dragon-slaying and ballroom dancing: the private supply of a public good}. Journal of Public Economics. vol. 25 pp. 1-12, 1984.
\bibitem{byers}
J. Byers, M. Mitzenmacher, and G. Zervas. \textit{Information Asymmetries  in Pay-Per-Bid Auctions, How Swoopo Makes Bank}.  arXiv:1001.0592, 2010.
\bibitem{mascolel}
A. Mas-colell, M. Whinston, and J. Green, \textit{Microeconomics Theory}.
Oxford, U.K., Oxford University Press, 1995.
\bibitem{hinosar}
T. Hinnosaar. \textit{Penny Auctions are Unpredictable}. Unpublished manuscript. Available at {http://toomas. hinnosaar.net/}, 2010.
\bibitem{kakhbod}
A. Kakhbod. \textit{Pay-to-bid auctions: To bid or not to bid}. Operations Research Letters,
vol. 41, no 5, pp. 462-467, 2013.
\bibitem{ned}
N. Augenblick. \textit{Consumer and Producer Behavior in the Market for Penny Auctions: A Theoretical and Empirical Analysis}. Unpublished manuscript. Available at {http://faculty.haas.berkeley.edu/ned/}, 2011. 
\bibitem{platt}
B. C. Platt, J. Price, and H. Tappen. \textit{The Role of Risk Preferences in Pay-to-Bid Auctions}. Management Science. To appear, 2013.
\bibitem{shubik}
M. Shubik. \textit{The Dollar auction game: A paradox in noncooperative
behavior and escalation}. Journal of Conflict Resolution. vol.15, no.1,  pp. 109-111, 1971.
\bibitem{Baye}
M. Baye,  D. Kovenock, and C. G. de Vries. \textit{The all-pay
auction with complete information}. Economic Theory, vol. 8, no. 2, pp. 291-305, 1996.
\bibitem{Smith}
J. M. Smith. \textit{The theory of games and the evolution of animal conflicts}. Journal of theoretical biology, vol. 47, no. 1, pp. 209-221, 1974.
\end{thebibliography}
\bibliographystyle{nonumber}

\newpage
\appendix
\bigskip

\noindent \textbf{\Large{Appendix}}\\


\textit{Notation}: \\
History of the game up to time $t, t\in \{1,2,3,\cdots\},$ which is common knowledge among the players,  is denoted by $h^t=(\vect{s}_1,\vect{s}_2,\cdots,\vect{s}_{t-1})$, where $\vect{s}_k=(s_{1,k},s_{2,k},\cdots,s_{n(t),k})$ represents the  strategy profile reported by the players in round $k$ of the game. Consider a sub-game beginning {after} history $h^t$. $\mathbb{E}_{i,t}[u(X_{h^t})]$ is the expected utility of player $i$ in this sub-game and, $\mathbb{E}_{i,t}[u(X_{h^t})|\{\mbox{Bid}\}]$ ($\mathbb{E}_{i,t}[u(X_{h^t})|\{\mbox{No Bid}\}]$) is the expected utility of player $i$ in this sub-game given that he plays $\{\mbox{Bid}\}$ ($\{\mbox{No Bid}\}$) in round $t$ and, $X_{h^t}$ is the random variable denoting money earned by the player in equilibrium in the sub-game starting after $h^t$. Each player at any round $t$ chooses a strategy  which is a map from any history of the game up to time $t$ to $[0,1]$,  a Bernoulli probability over $\{\mbox{Bid, No Bid}\}$. Let $p_t(h^t), 0\leq p_t(h^t)\leq 1,$ be the probability of choosing $\{\mbox{Bid}\}$ after observing history $h^t$ of the game. Number of players present  in round $t, t\in \{1,2,3,\cdots\}$ of the game is denoted by $n_t$. Note that when re-entry is allowed $n_t=n$   in any round $t, t\in \{1,2,3,\cdots\}$.\\


\begin{proof}[\textbf{{Proof of Theorem \ref{th1}}}]
We prove each part of the theorem, separately, as follows.\\
\noindent{\textbf{Proof of I.}}:\\
 We note that in any SSPE,  at any round $t$, $0<p_t(h^t)<1$, because, due to the game specification, for example if $p_t(h^t)=1$ then for any player it is profitable to unilaterally deviate to $\{\mbox{No Bid}\}$ (similar argument holds when $p_t(h^t)=0$). Further, it can be shown that there exists $\delta$ such that at any round $t$ of the game $p_t(h^t)>\delta$. We prove this by contradiction. Suppose that there is not such $\delta$, ie, for any $\delta>0$ there exists history $h^t$, which occurs with positive probability in equilibrium, such that $p_t(h^t)<\delta$. Thus, by picking $\delta$ sufficiently small we have:
\begin{eqnarray}
\mathbb{E}_{i,t}[u(X_{h^t})|\{\mbox{Bid}\}]& \stackrel{(a)}{>}&(1-\delta)^{n-1}u(v-s+w_i-c)\nonumber\\
&\geq& \frac{u(v)}{n}\nonumber \\&\geq& \mathbb{E}_{i,t}[u(X_{h^t})]\nonumber\\
& \geq & \mathbb{E}_{i,t}[u(X_{h^t})|\{\mbox{No Bid}\}],
\end{eqnarray}
where $(a)$ follows since $\delta$ is very small. \\
Next, we show that in any sub-game perfect equilibrium starting in round $t$ of the game  each player earns expected utility $u(w)$ where $w$ is the the wealth of the player in round $t$.\\
First, we show this statement is true when $w=0$, that is, $\mathbb{E}_{i,t}[u(X_{h^t})]=0$ when $w=0$ in round $t$. We prove it by contradiction. Suppose that  there exists history $h^t$, which occurs with positive probability in equilibrium, such that
\begin{eqnarray}
\mathbb{E}_{i,t}[u(X_{h^t})]=\gamma>0
\end{eqnarray}
As we proved in the above, there exists $\delta$ such that $p_t(h^t)>\delta, \forall \ t$. Further,  since, $p_t(h^t)\in (0,1), \forall t,$ each player should be indifferent between choosing $\{\mbox{Bid}\}$ and $\{\mbox{No Bid}\}$, in each round $t$. Also, since $\mathbb{E}_{i,t}[u(X_{h^t})]=\gamma>0$, there exists $h^s$ including $h^t$, ie, $h^t\subset h^s$, such that $\mathbb{E}_{i,t}[u(X_{h^s})]\geq \gamma$. 
Thus, due to the fact player $i$ indifferent between $\{\mbox{Bid}\}$ and $\{\mbox{No Bid}\}$ we have:
\begin{eqnarray}
\label{11}
\mathbb{E}_{i,s}[u(X_{h^s})]&=&\mathbb{E}_{i,s}[u(X_{h^s})|\{\mbox{No Bid}\}]\nonumber\\
&=&\Lambda_{i,s} \ \mathbb{E}_{i,s+1}[u(X_{h^{s+1}})]\geq \gamma \nonumber\\
&&\Rightarrow \quad \mathbb{E}_{i,s}[u(X_{h^{s+1}})] \geq \frac{\gamma}{\Lambda_{i,s}},
\end{eqnarray}
where, $\Lambda_{i,s}$ is the probability that at least two players (except $i$) play $\{\mbox{Bid}\}$ in round $s,$ given $h^s$, ie, 
\begin{eqnarray*}
\Lambda_{i,s}:= 1- \binom{n_s-1}{1}p_s(h^s)(1-p_s(h^s))^{n-2},
\end{eqnarray*}
where $n_s$ is the number of players in round $s$, (note that when re-entry is allowed $n_s=n, \ \forall s$.)
Furthermore, note that $0<\Lambda_{i,s}<1.$ Using the new lower  bound  derived in \eqref{11} and, following  similar the arguments we did to derive \eqref{11} imply
\begin{eqnarray}
\mathbb{E}_{i,s+2}[u(X_{h^{s+2}})] \geq \frac{\gamma}{\Lambda_{i,s}\Lambda_{i,s+1}}, \quad h^{s+2}\subset h^{s+1}\subset h^s.
\end{eqnarray}
Following the above arguments we obtain
\begin{eqnarray}
\mathbb{E}_{i,s+r}[u(X_{h^{s+r}})] \geq \frac{\gamma}{\prod_{k=0}^{r-1}\Lambda_{i,s+k}},
\end{eqnarray}
where $h^{s+r}\subset \cdots \subset h^{s+1}\subset h^s$.
But notice since for any $k\geq 0$,   $0<\Lambda_{i,s+k}<1$, then by choosing $r$ sufficiently large we can get 
\begin{eqnarray}
 \frac{\gamma}{\prod_{k=0}^{r-1}\Lambda_{i,s+k}} > u(v),
\end{eqnarray}
that is a contradiction, since $\mathbb{E}_{i,s+r}[u(X_{h^{s+r}})]$ can not exceed $u(v)$.  Therefore, 
 \begin{eqnarray}
 \label{22}
\mathbb{E}_{i,t}[u(X_{h^t})]=0.
\end{eqnarray}
Now, suppose that $w>0$, then Eq. \eqref{22} along with \eqref{ut} imply
\begin{align}
\label{222}
\mathbb{E}_{i,t}[u(w+X_{h^t})]&=\mathbb{E}_{i,t} \left[\frac{1-e^{-\rho(w+X_{h^t})}}{\rho}\right] \nonumber\\
&= \frac{1- \mathbb{E}_{i,t}\left[e^{-\rho(w+X_{h^t})}\right]}{\rho} \nonumber\\
&= \frac{1- e^{-\rho w}\mathbb{E}_{i,t}\left[e^{-\rho X_{h^t}}\right]}{\rho} \nonumber\\
&= \frac{1- e^{-\rho w}+e^{-\rho w}\left(1-\mathbb{E}_{i,t}\left[e^{-\rho X_{h^t}}\right]\right)}{\rho} \nonumber\\
&=  \frac{1- e^{-\rho w}}{\rho}+e^{-\rho w}\mathbb{E}_{i,t} \left[\frac{1-e^{-\rho X_{h^t}}}{\rho}\right]  \nonumber\\
&=  \frac{1- e^{-\rho w}}{\rho}+e^{-\rho w}\mathbb{E}_{i,t}[u(X_{h^t})]\nonumber\\
&= u(w).
\end{align}
The above equality completes the proof of the first part of the theorem. Further, note that \eqref{222} follows whether re-entry is allowed or not. \\

Before proving the next part,   we present the following Remark that holds in risk-loving utility function.\begin{Remark}[see \cite{mascolel}]
\label{Remark}
Let $u(\cdot)$ be a  CARL utility function. Suppose $X$ and $Y$ are two random variables and $\alpha$ is a constant. Then, 
\begin{eqnarray*}
\mathbb{E}[u(X+\alpha)] \geq \mathbb{E}[u(Y+\alpha)]  \quad   \Longleftrightarrow \quad  \mathbb{E}[u(X)] \geq \mathbb{E}[u(Y)] 
\end{eqnarray*}
\end{Remark}
 Remark \ref{Remark} states that comparing two random variables is independent of a constant shift when the utility functions has a CARL form, like what we defined in \eqref{ut}.\\

\noindent{\textbf{Proof of II. (when re-entry is allowed)}}:\\
The expected utility  of player $i$ if he plays $\{\mbox{Bid}\}$ (with wealth level equal to $w_{i,t}$ at the begging of round $t$) in equilibrium in the sub-game starting after history  $h^t$  is equal to,
\begin{align}
\label{33}
&\mathbb{E}_{i,t}[u(w_{i,t}+X_{h^t})|\{\mbox{Bid}\}]=  u(w_{i,t}+v-s-c) (1-p_t(h^t))^{n-1}   \nonumber\\  
& \quad  \quad+\sum_{k=1}^{n-1}\Bigg[\binom{n-1}{k}  p_t(h^t)^{k}(1-p_t(h^t))^{n-1-k}\times 
\mathbb{E}_{i,t+1}[u(w_{i,t}-c+X_{h^t\cup \vect{s}_{t+1}^{k}})]\Bigg],
\end{align}

where   $\vect{s}_{t+1}^{k}$ is the strategy profile in round $t+1$ that $k$ players play $\{\mbox{Bid}\}$ and $h^t\cup \vect{s}_{t+1}^{k}$ is the updated history. In \eqref{33}, the first term corresponds to the event that player $i$ wins the object, ie, the event in which player $i$ is the only player who plays $\{\mbox{Bid}\}$ in round $t$  and, the second term corresponds to the expected utility of player $i$ from continuing after history $h^t$, equivalently,  more than one player play $\{\mbox{Bid}\}$. \\
Similarly, if player $i$ plays $\{\mbox{No Bid}\}$, then 
\begin{align}
\label{44}
&\mathbb{E}_{i,t}[u(w_{i,t}+X_{h^t})|\{\mbox{No Bid}\}]=  u(w_{i,t})\binom{n-1}{1}(1-p_t(h^t))^{n-2}p_t(h^t)       \nonumber\\
& \quad \quad \quad+\sum_{k\neq 1}\Bigg[\binom{n-1}{k}  p_t(h^t)^{k}(1-p_t(h^t))^{n-1-k} \times  \mathbb{E}_{i,t+1}[u(w_{i,t}+X_{h^t\cup \vect{s}_{t+1}^{k}})]\Bigg]
\end{align}
In \eqref{44}, the first term corresponds to the event that player $j$ ($j\neq i$) wins  the object, ie,  the event in which player $j$ ($j\neq i$) is the only player who plays $\{\mbox{Bid}\}$ in round $t$  and, the second term corresponds to the expected utility of player $i$ from continuing after history $h^t$, equivalently,  more than one player other than $i$ play $\{\mbox{Bid}\}$. \\

Since player $i$ purely randomizes between $\{\mbox{Bid}\}$ and $\{\mbox{No Bid}\}$, then  player $i$ is indifferent between choosing $\{\mbox{Bid}\}$ and $\{\mbox{No Bid}\}$, that is, \eqref{33} is equal to \eqref{44}.\\  

Now, using  Remark \eqref{Remark} enable us to simplify  \eqref{33} and \eqref{44} as follows. \\
Without loss of generality, we can set $w_{i,t}=c$ and, consequently,   obtain
\begin{align}
\label{444}
\mathbb{E}_{i,t+1}[u(w_{i,t}-c+X_{h^t\cup \vect{s}_{t+1}^{k}})]\big|_{w_{i,t}=c}&=  \mathbb{E}_{i,t+1}[u(X_{h^t\cup \vect{s}_{t+1}^{k}})]\nonumber\\
&\stackrel{(a)}{=}0.
\end{align}
\begin{eqnarray}
\label{4444}
 \mathbb{E}_{i,t+1}[u(w_{i,t}+X_{h^t\cup \vect{s}_{t+1}^{k}})]\big|_{w_i=c}&=& \mathbb{E}_{i,t+1}[u(c+X_{h^t\cup \vect{s}_{t+1}^{k}})]\nonumber\\
 &\stackrel{(b)}{=}& u(c),
\end{eqnarray}
where $(a)$ follows from \eqref{22}, and  $(b)$ from \eqref{222}. \\
Now, plugging  $w_i=c$ into \eqref{33} (because of Remark \ref{Remark}) and using \eqref{444}, Eqs. \eqref{33}   can be simplified as follows
\begin{align}
\label{333}
\mathbb{E}_{i,t}[u(w_{i,t}+X_{h^t})|\{\mbox{Bid}\}]\big|_{w_{i,t}=c}=u(v-s) (1-p_t(h^t))^{n-1}  
\end{align}
and similarly,  plugging  $w_{i,t}=c$ into \eqref{44}  and employing  \eqref{4444}, then \eqref{44}  is  simplified as follows 
\begin{align}
\mathbb{E}_{i,t}[u(w_{i,t}+X_{h^t})|\{\mbox{No Bid}\}]\big|_{w_{i,t}=c}&=u(c)\binom{n-1}{1}(1-p_t(h^t))^{n-2}p_t(h^t)  \nonumber \\ 
&  \quad + \sum_{k \neq 1}^{n-1}\binom{n-1}{k}  p_t(h^t)^{k}(1-p_t(h^t))^{n-1-k}u(c) \nonumber\\
&= u(c) \Bigg[\binom{n-1}{1}(1-p_t(h^t))^{n-2}p_t(h^t) \nonumber\\
& \quad + \sum_{k \neq 1}^{n-1}\binom{n-1}{k}  p_t(h^t)^{k}(1-p_t(h^t))^{n-1-k} \Bigg] \nonumber\\
\label{3333}
&= u(c).
\end{align}
Finally, since player $i$   is indifferent between choosing $\{\mbox{Bid}\}$ and $\{\mbox{No Bid}\}$, then equating \eqref{333} and \eqref{3333} gives that
\begin{eqnarray}
\label{mohem}
p_t(h^t)=1- \sqrt[n-1]{\frac{u(c)}{u(v-s)}}.
\end{eqnarray}
Now, it is immediate from \eqref{mohem} that the characterized  symmetric equilibrium  is  unique and of course stationary since it is controlled by the  $(n, v, s, c)$. 
\\

\noindent{\textbf{Proof of III. (when re-entry is \emph{not} allowed)}}:\\
Let player $i$ be one of $n_t$ (remaining) players  who are left  to play in round $t$ of the game. Then,
the expected utility  of player $i$ if he plays $\{\mbox{Bid}\}$ (with wealth level equal to $w_{i,t}$ at the begging of round $t$) in equilibrium in the sub-game starting after history  $h^t$  is equal to,
\begin{align}
\label{33-1}
&\mathbb{E}_{i,t}[u(w_{i,t}+X_{h^t})|\{\mbox{Bid}\}]=  u(w_i+v-s-c) (1-p_t(h^t))^{n_t-1}   \nonumber\\  
& \quad  \quad+\sum_{k=1}^{n_t-1}\Bigg[\binom{n_t-1}{k}  p_t(h^t)^{k}(1-p_t(h^t))^{n_t-1-k}\times 
\mathbb{E}_{i,t+1}[u(w_{i,t}-c+X_{h^t\cup \vect{s}_{t+1}^{k}})]\Bigg],
\end{align}

where   $\vect{s}_{t+1}^{k}$ is the strategy profile in round $t+1$ that $k$ players of  the remaining  $n_t$ players of round $t$,  play $\{\mbox{Bid}\}$ and $h^t\cup \vect{s}_{t+1}^{k}$ is the updated history. In \eqref{33-1}, the first term corresponds to the event that player $i$ wins the object, ie, the event in which player $i$ is the only player (among the remaining players) who plays $\{\mbox{Bid}\}$ in round $t$  and, the second term corresponds to the expected utility of player $i$ from continuing after history $h^t$, equivalently,  more than one player play $\{\mbox{Bid}\}$. \\
If player $i$ plays $\{\mbox{No Bid}\}$, then 
\begin{align}
\label{44-1}
\mathbb{E}_{i,t}[u(w_{i,t}+X_{h^t})|\{\mbox{No Bid}\}]&=  u(w_{i,t}) \left[\binom{n_t-1}{1}(1-p_t(h^t))^{n_t-2}p_t(h^t)\right]    \nonumber \\   
&\quad+u(w_{i,t}) \left[\sum_{k\geq 2}\binom{n_t-k}{k}(1-p_t(h^t))^{n_t-k}p_t(h^t)^k\right]  \nonumber \\   
&\quad+ \mathbb{E}_{i,t+1}[u(w_{i,t}+X_{h^t\cup \vect{s}_{t+1}^{k}})] (1-p_t(h^t))^{n_t-1}.
\end{align}
In \eqref{44-1}, the first term corresponds to the event that player $j$ ($j\neq i$) wins  the object, ie,  the event in which player $j$, one of the remaining players  ($j\neq i$),  is the \emph{only} player who plays $\{\mbox{Bid}\}$ in round $t$.  The second term is corresponding to the case in which more than two players among the remaining ones play $\{\mbox{Bid}\}$ in round $t$. Moreover, note that since re-entry is not allowed,  the first two terms of \eqref{44-1} represent events in which player $i$   will be out from the rest of the game.
The third term of \eqref{44-1} represents the \emph{only} event in which player $i$ maintains in the game with playing  $\{\mbox{No Bid}\}$, that is the tie breaking case.
Since re-entry is not allowed, by playing $\{\mbox{No Bid}\}$ player $i$ remains in the game \emph{only if all}
the other remaining players play $\{\mbox{No Bid}\}$ as well. This event is captured by the last term of 
\eqref{44-1}.\\

Now, since player $i$ purely randomizes between $\{\mbox{Bid}\}$ and $\{\mbox{No Bid}\}$, then  player $i$ is indifferent between choosing $\{\mbox{Bid}\}$ and $\{\mbox{No Bid}\}$, that is, \eqref{33-1} is equal to \eqref{44-1}.

Again, similar to the case where re-entry is permitted,  by using  Remark \eqref{Remark} and setting  $w_i=c$ and, we  obtain
\begin{align}
\label{444-1}
\mathbb{E}_{i,t+1}[u(w_{i,t}-c+X_{h^t\cup \vect{s}_{t+1}^{k}})]\big|_{w_{i,t}=c}&=  \mathbb{E}_{i,t+1}[u(X_{h^t\cup \vect{s}_{t+1}^{k}})]\nonumber\\
&\stackrel{(a)}{=}0.
\end{align}
\begin{eqnarray}
\label{4444-1}
 \mathbb{E}_{i,t+1}[u(w_{i,t}+X_{h^t\cup \vect{s}_{t+1}^{k}})]\big|_{w_{i,t}=c}&=& \mathbb{E}_{i,t+1}[u(c+X_{h^t\cup \vect{s}_{t+1}^{k}})]\nonumber\\
 &\stackrel{(b)}{=}& u(c),
\end{eqnarray}
where $(a)$ follows from \eqref{22}, and  $(b)$ from \eqref{222}. Now, plugging  $w_{i,t}=c$ into \eqref{33-1} and using \eqref{444-1}, Eqs. \eqref{33-1}   is  simplified as follows
\begin{align}
\label{333-1}
\mathbb{E}_{i,t}[u(w_{i,t}+X_{h^t})|\{\mbox{Bid}\}]\big|_{w_{i,t}=c}=u(v-s) (1-p_t(h^t))^{n-1}  
\end{align}
and similarly,  plugging  $w_{i,t}=c$ into \eqref{44-1}  and using  \eqref{4444-1}, then  \eqref{44-1}  is  simplified as follows 
\begin{align}\label{3333-1}
\mathbb{E}_{i,t}[u(w_{i,t}+X_{h^t})|\{\mbox{No Bid}\}]\big|_{w_i=c}&= u(c) \left[\binom{n_t-1}{1}(1-p_t(h^t))^{n_t-2}p_t(h^t)\right]    \nonumber \\   
&\quad+u(c) \left[\sum_{k\geq 2}\binom{n_t-k}{k}(1-p_t(h^t))^{n_t-k}p_t(h^t)^k\right]  \nonumber \\   
&\quad+ \mathbb{E}_{i,t+1}[u(c+X_{h^t\cup \vect{s}_{t+1}^{k}})] (1-p_t(h^t))^{n_t-2} \nonumber\\
&=u(c) \left[\sum_{k\geq 1}\binom{n_t-k}{k}(1-p_t(h^t))^{n_t-k}p_t(h^t)^k\right]  \nonumber \\   
&\quad+ u(c)(1-p_t(h^t))^{n_t-1} \nonumber\\
&= u(c).
\end{align}

Moreover, since player $i$   is indifferent between choosing $\{\mbox{Bid}\}$ and $\{\mbox{No Bid}\}$, then equating \eqref{333-1} and \eqref{3333-1} gives that
\begin{eqnarray}
\label{mohem-1}
p_t(h^t)=1- \sqrt[n_t-1]{\frac{u(c)}{u(v-s)}}.
\end{eqnarray}
Equation \eqref{mohem-1} reveals that the symmetric equilibrium is unique.  Also the equilibrium  strategy that each player follows in each round $t$ of the game depends only on the relevant state variables $(n_t, v, s, c)$, therefore it is Markov perfect.

\end{proof}


\begin{proof}[\textbf{{Proof of Theorem \ref{th3}}}]
The probability of the event that  the game ends in round $t$ of the game given that $t$ is reached is equal to
\begin{eqnarray}
h_t=\frac{n_tp_t(1-p_t)^{n_t-1}}{1-(1-p_t)^{n_t}}
\end{eqnarray}
 $h_t$ is called \emph{hazard rate} at time $t$.
\\
The probability thst a round $t$ is reached is the equal to the probability the game doesn't  end in any round $s, s<t$, which is equal to
\begin{eqnarray}
\prod_{s=1}^{t-1}(1-h_s)=\prod_{s=1}^{t-1}\left(1-\frac{np_s(1-p_s)^{n_s-1}}{1-(1-p_s)^{n_s}}\right).
\end{eqnarray}
Now, let $Q_{n_t,t}$ denote the expected number of entrants in round $t$  when the number remaining players is  $n_t$. Thus,
\begin{eqnarray}
\label{19}
Q_{n_t,t}&=& \sum_{k=1}^{n_t}\binom{n_t}{k}p_t^k(1-p_t)^{n_t-k}k+(1-p_t)^{n_t}Q_{n_t,t}\nonumber\\
&=&\sum_{k=0}^{n}\binom{n_t}{k}p_t^k(1-p_t)^{n_t-k}k+(1-p_t)^{n_t}Q_{n_t,t}\nonumber\\
&=&np_t+(1-p_t)^{n_t}Q_{n_t,t}
\end{eqnarray}
Equation \eqref{19} implies that
\begin{eqnarray}
Q_{n_t,t}= \frac{n_tp_t}{1-(1-p_t)^{n_t}}.
\end{eqnarray}
The seller's expected earnings from the bid fees in round $t$ is the expected number of entrants times the bid fee times the probability that round $t$ is reached, that is 
\begin{eqnarray*}
 Q_{n_t,t}\times c \times \prod_{s=1}^{t-1}(1-h_s)=\frac{n_tp_t}{1-(1-p_t)^{n_t}} \times c \times \prod_{s=1}^{t-1}\left(1-\frac{np_s(1-p_s)^{n_s-1}}{1-(1-p_s)^{n_s}}\right).
\end{eqnarray*}
Therefore, the seller's expected earning from the bid fees throughout the game is exactly equal to 
\begin{eqnarray*}
 \sum_{t=1}^{\infty}\left[Q_{n_t,t}\times c \times \prod_{s=1}^{t-1}(1-h_s)\right]= \sum_{t=1}^{\infty}\left[\frac{n_tp_t}{1-(1-p_t)^{n_t}} \times c \times \prod_{s=1}^{t-1}\left(1-\frac{n_sp_s(1-p_s)^{n_s-1}}{1-(1-p_s)^{n_s}}\right)\right].
\end{eqnarray*}
\end{proof}


\begin{proof}[\textbf{{Proof of Theorem \ref{th333}}}]
Keep the bidding fee fixed. As we proved in Theorem \ref{th2}, in equilibrium, in any round $t$, the probability that a player exits (plays $\{\mbox{No Bid}\}$) when there are $n_t$ players remained in the game is uniquely determined as a function of the objects's value $v$, the bid fee $c$, and the sale price $s$ as 
\begin{eqnarray*}
p_t= 1-\sqrt[n_t-1]{\frac{u(c)}{u(v-s)}}.
\end{eqnarray*}
When $\frac{v-s}{c} \rightarrow \infty$, $\frac{u(c)}{u(v-s)} \rightarrow 0$ and therefore $p_t \rightarrow 1.$
Notice that, when re-entry is permitted the above probability becomes stationary because $n_t=n$, for all $t$. Moreover, as shown in the previous theorem, the seller's expected profit (earnings from the bid fees) in round $t$ is given by
\begin{eqnarray*}
 Q_{n_t,t}\times c \times \prod_{s=1}^{t-1}(1-h_s)=\frac{n_tp_t}{1-(1-p_t)^{n_t}} \times c \times \prod_{s=1}^{t-1}\left(1-\frac{np_s(1-p_s)^{n_s-1}}{1-(1-p_s)^{n_s}}\right).
\end{eqnarray*}
 Since $c$ is fixed, and $p_t \rightarrow 1$, it follows that the above expression is maximized. Consequently, the overall seller's profit is maximized.

\end{proof}


\begin{proof}[\textbf{{Proof of Theorem \ref{th2}}}] We prove each part of the theorem separately as follows.\\
In the following we first prove the second part of the theorem and then the rest. \\
\noindent{\textbf{Proof of II.}}:\\
As we proved in Theorem \ref{th1}, when re-entry is allowed,  in equilibrium, in each stage of the game, each player chooses to play $\{\mbox{Bid}\}$ with following \emph{stationary} probability
	\begin{eqnarray}
	\label{l1}
	p=1- \sqrt[n-1]{\frac{u(c)}{u(v-s)}}. 
	\end{eqnarray}
The stationarity comes from the fact that the above probability  is  controlled by the constant parameters of the model that are $n, s, v,$ and $c$. \\
Moreover, as we proved in Theorem \ref{th3},  the seller's expected earning \emph{only} from the bid fees throughout the game is exactly equal to 
\begin{eqnarray*}
 \sum_{t=1}^{\infty}\left[\frac{np_t}{1-(1-p_t)^n} \times c \times \prod_{s=1}^{t-1}\left(1-\frac{np_s(1-p_s)^{n-1}}{1-(1-p_s)^{n}}\right)\right].
\end{eqnarray*} 

Now, due to the fact that when re-entry is allowed the equilibrium is stationary (ie, because of \eqref{l1}, $p_t=p$ for any $t\in\{1,2,3,\cdots\}$), the above quality can be simplified as follows. \\

Since the seller's expected revenue is equal to the sale price, ie, $s$,  plus the seller's expected earning  from the bid fees throughout the game, then we obtain

\begin{eqnarray} \label{revenue}
 \mbox{Seller's expected revenue}&=&{s}+\sum_{t=1}^{\infty}\left[\left(c\frac{np}{1-(1-p)^{n}}\right)\left[1-\frac{np(1-p)^{n-1}}{1-(1-p)^{n}}\right]^{t-1}\right]\nonumber\\
&=& s+ \left(c\frac{np}{1-(1-p)^{n}}\right) \sum_{t=1}^{\infty} \left[1-\frac{np(1-p)^{n-1}}{1-(1-p)^{n}}\right]^{t-1}\nonumber \\
&=& s+\left(c\frac{np}{1-(1-p)^{n}}\right) \frac{1-(1-p)^{n}}{np(1-p)^{n-1}}\nonumber\\
&=& s+\frac{c}{(1-p)^{n-1}} \nonumber \\
&\stackrel{(a)}{=}&s+  c \frac{u(v-s)}{u(c)}.
\end{eqnarray} 

\noindent{\textbf{Proof of I.}}:\\
When players are risk neutral (ie, $u(x)=x$),  then \eqref{revenue} implies that
\begin{eqnarray}
\label{revenuen2}
\mbox{Seller's expected revenue (with risk neutral utility)}=v.
\end{eqnarray}
\noindent{\textbf{Proof of III.}}:\\
It is immediate from \eqref{revenue} that the seller's expected revenue is independent of number of players.
\\
\noindent{\textbf{Proof of IV.}}:\\
Now, we show that when players are risk-loving, ie, $\rho \neq 0$, 
\begin{eqnarray}
\label{b1}
s+  c \frac{u(v-s)}{u(c)} > v.
\end{eqnarray}
To prove \eqref{b1} we use the following Lemma.
\begin{Lemma}
\label{lem}
Let $u(\cdot)$ be a convex function and $x>0$ is a constant. Define
$$
f(\alpha):=u((1+\alpha)x)-(1+\alpha)u(x).
$$
Then, $f(\alpha)>0$.
\end{Lemma}
\begin{proof}[Proof of Lemma \ref{lem}]
To prove Lemma \ref{lem}, we first show that
\begin{eqnarray}
\label{25}
&&\frac{u(x)}{x}=\frac{\int_{0}^x u'(t)dt}{x}\stackrel{(a)}{<}\frac{\int_{0}^x u'(x)dt}{x}=u'(x) \nonumber \\
&& \quad \quad \Rightarrow  \quad u(x)<x u'(x).
\end{eqnarray}
where $(a)$ follows because $u(\cdot)$ is increasing.\\
Now, using \eqref{25} we obtain
\begin{eqnarray}
f'(\alpha)=xu'((1+\alpha)x)-u(x)>xu'(x)-u(x)>0.
\end{eqnarray}
Thus $f(\alpha)$ is increasing in $\alpha$ and $f(\alpha)>f(0)=0$.
\end{proof}

As a consequence of  Lemma \ref{lem} we have
\begin{eqnarray}
\label{temp}
x>y \quad \Rightarrow \quad \frac{u(x)}{u(y)}>\frac{x}{y}.
\end{eqnarray}
because we can simply set $x=(1+\alpha)y$, where $\alpha>0$.
\\
Now, we can prove  \eqref{b1}  holds as follows
\begin{eqnarray}
s+  c \frac{u(v-s)}{u(c)} - v= c\left[ \frac{u(v-s)}{u(c)}-\frac{v-s}{c}\right]\stackrel{(a)}{>}0
\end{eqnarray}
where $(a)$ is correct because  $v-s>c$ and \eqref{temp}.

\noindent{\textbf{Proof of V.}}:\\
In the following, we show that the seller's revenue is increasing in $v$, decreasing in $s$ and decreasing in $c$.
\begin{eqnarray}
&&\frac{\partial \left[s+  c \frac{u(v-s)}{u(c)}\right]}{\partial v}>0. \nonumber\\
&&\frac{\partial \left[s+  c \frac{u(v-s)}{u(c)}\right]}{\partial s}=1-c\frac{u'(v-s)}{u(c)}<1-c\frac{u'(c)}{u(c)}\stackrel{(a)}{<}0. \nonumber\\
&&\frac{\partial \left[s+  c \frac{u(v-s)}{u(c)}\right]}{\partial c}=\frac{u(v-s)u(c)-cu'(c)u(v-s)}{u(c)^2} =\frac{u(v-s)}{u(c)^2}(u(c)-cu'(c))\stackrel{(b)}{<}0 \nonumber \\
\end{eqnarray}
where $(a)$ and $(b)$ are both followed by \eqref{25}.\\
\noindent{\textbf{Proof of VI.}}:\\
Because of the results of  the previous part, the expected revenue of the seller  is decreasing in $s$ and decreasing in $c$.  Therefore, in order to find the maximum value of the seller's expected  revenue (when $v$ is given), in \eqref{revenue}, we set $s=0$ and take a limit when $c\rightarrow 0$ as follows:
\begin{eqnarray}
\max_{s,c}\left[s+  c \frac{u(v-s)}{u(c)}\right]=\lim_{c \rightarrow 0} \frac{cu(v)}{u(c)}=\frac{u(v)}{u'(0)}=\frac{1-e^{-\rho v}}{\rho}=u(v).
\end{eqnarray}

\noindent{\textbf{Proof of VII.}}:\\
Consider a seller offering lottery tickets for a prize/object of (monetary) value $v$ to $n$ potential buyers with CARL utility. Cost of buying a ticket for each player is equal to $c>0$, and the winner is chosen randomly. Thus, the maximal price the seller can choose to charge for a (lottery) ticket is the solution of $u(c)=\frac{1}{n}u(v)$. Hence, the optimal (revenue maximizing) ticket price is $c^*=u^{-1}\left(\frac{1}{n}u(v)\right)$.  Since for all $x>0$, $u(x)>x$, thus, for all $x>0$,  $x>u^{-1}(x)$. Therefore, 
$\frac{1}{n}u(v)>u^{-1}\left(\frac{1}{n}u(v)\right)$, and consequently for all $n\in \mathbb{N}$, $u(v)>n\times u^{-1}\left(\frac{1}{n}u(v)\right)=nc^*=\mbox{the maximum seller's revenue from the lottery}$.

As we have shown, in the previous part, Part (VI), $u(v)$ is the maximum seller's revenue in the game proposed in Section \ref{sec1}. Thus,  the above inequality implies that,   for sufficiently small biding fee,  the seller's revenue is strictly \emph{higher} than a standard lottery, for any number of players.

\end{proof}
%

\begin{proof}[\textbf{{Proof of Theorem \ref{th4}}}]
We prove each part of the theorem separately as follows.\\
\noindent{\textbf{Proof of I.}}:\\
As we proved in Theorem \ref{th2}, in equilibrium, the probability that a player exits (plays $\{\mbox{No Bid}\}$) when there are $n$ players remainng in the game is uniquely determined as a function of the objects's value $v$, the bid fee $c$, and the sale price $s$ as 
\begin{eqnarray}
q:=1- \left(1-\sqrt[n-1]{\frac{u(c)}{u(v-s)}}\right)=\sqrt[n-1]{\frac{u(c)}{u(v-s)}}.
\end{eqnarray}
When $\frac{v-s}{c} \rightarrow \infty$, $\frac{u(c)}{u(v-s)} \rightarrow 0$ and therefore $q \rightarrow 0.$ For simplicity of exposition, we denote $\frac{u(c)}{u(v-s)}$ by $\lambda$ and write
$$
q_n(\lambda)=\sqrt[n-1]{\lambda} \quad \Rightarrow \quad \lim_{\lambda\rightarrow 0}q_n(\lambda)=0.
$$

 Let ${Z}_{n,\lambda}$ be a random variable denoting the number of bids (number of players playing $\{\mbox{Bid}\}$)  in a round without entry if there are $n$ remaining players and each chooses to play $\{\mbox{No Bid}\}$ with probability $q_n(\lambda)$. Thus,
\begin{eqnarray}
\mbox{Prob}\{Z_{n,\lambda}=m\}=\frac{\binom{n}{m}(1-q_n(\lambda))^mq_n(\lambda)^m}{1-(q_n(\lambda))^n}, \quad \forall m\in \{1,2,\cdots,n\}.
\end{eqnarray}
Thus, the probability the auction ends in any particular round, given that it is reached and has $n$ remaining players, is 
\begin{eqnarray}
\mbox{Prob}\{Z_{n,\lambda}=1\}=\frac{n(1-q_n(\lambda)(q_n(\lambda))^{n-1}}{1-(q_n(\lambda))^n}.
\end{eqnarray}
Since as $\lambda \rightarrow 0$, $q_n(\lambda)\rightarrow 0$, then we have 
\begin{eqnarray}
\label{tem11}
\lim_{\lambda \rightarrow 0} \mbox{Prob}\{Z_{n,\lambda}=1\}=\frac{n(1-q_n(\lambda)(q_n(\lambda))^{n-1}}{1-(q_n(\lambda))^n}=0.
\end{eqnarray}
Since \eqref{tem11} holds for any $n$,  it follows  that the expected length of the auction tends to infinity.
\\
 
\noindent{\textbf{Proof of II.}}:\\
Now we prove the second part of the theorem. It follows  that
\begin{eqnarray}
\label{tem1}
\lim_{\lambda \rightarrow 0} \frac{\mbox{Prob}\{Z_{n,\lambda}=m\}}{\mbox{Prob}\{Z_{n,\lambda}=m+1\}}=\lim_{\lambda \rightarrow 0} \frac{(m+1)q_n(\lambda)}{(n-1)(1-q_n(\lambda))}=0.
\end{eqnarray}
Equation \eqref{temp} implies that, while, $\mbox{Prob}\{Z_{n,\lambda}=2\}$ tends to $0$, it is an order of magnitude larger than $\mbox{Prob}\{Z_{n,\lambda}=1\}$ when $\lambda$ is small. Now, let the random variable $T_{n,m}$ be  the number of rounds without re-entry before an auction with $n$ players is reduced to one with no more than $m$ remaining players. Therefore, $T_{n,1}$ means the last round of the game and $T_{n,m} < T_{n,1}$ implies that there is some round with fewer than $m$ players before the end of the auction. We can write
\begin{eqnarray}
\mbox{Prob}\{T_{n,2}<T_{n,1}\}&=&\mbox{Prob}\{Z_{n,\lambda}=2\}+\sum_{m=3}^n\mbox{Prob}\{Z_{n,\lambda}=m\}\mbox{Prob}\{T_{m,2}<T_{m,1}\} \nonumber \\ \label{t2}
&=&\mbox{Prob}\{Z_{n,\lambda}=2\}+\sum_{m=3}^{n-1}\mbox{Prob}\{Z_{n,\lambda}=m\}\mbox{Prob}\{T_{m,2}<T_{m,1}\}\nonumber\\
&&+\mbox{Prob}\{Z_{n,\lambda}=n\}\mbox{Prob}\{T_{n,2}<T_{n,1}\}
\end{eqnarray}
Equation \eqref{t2} implies that
\begin{eqnarray}\label{tem3}
\mbox{Prob}\{T_{n,2}<T_{n,1}\}=\frac{\mbox{Prob}\{Z_{n,\lambda}=2\}+\sum_{m=3}^{n-1}\mbox{Prob}\{Z_{n,\lambda}=m\}\mbox{Prob}\{T_{m,2}<T_{m,1}\}}{1-\mbox{Prob}\{Z_{n,\lambda}=n\}}
\end{eqnarray}
Now, we show that \eqref{tem3} tends to 1 as $\lambda$ tends to 0. We show this by induction over $n$. First, consider the case where $n=3$. Then \eqref{tem3} is simplified as follows
\begin{eqnarray}\label{t4}
\mbox{Prob}\{T_{3,2}<T_{3,1}\}=\frac{\mbox{Prob}\{Z_{3,\lambda}=2\}}{1-\mbox{Prob}\{Z_{3,\lambda}=3\}}=\frac{\mbox{Prob}\{Z_{3,\lambda}=2\}}{\mbox{Prob}\{Z_{3,\lambda}=2\}+\mbox{Prob}\{Z_{3,\lambda}=1\}}.
\end{eqnarray}
 
Equation \eqref{tem1} along with \eqref{t4} imply that
 
\begin{eqnarray}
\lim_{\lambda \rightarrow 0}\frac{\mbox{Prob}\{Z_{3,\lambda}=2\}+\mbox{Prob}\{Z_{3,\lambda}=1\}}{\mbox{Prob}\{Z_{3,\lambda}=2\}}=1+\lim_{\lambda \rightarrow 0}\frac{\mbox{Prob}\{Z_{3,\lambda}=1\}}{\mbox{Prob}\{Z_{3,\lambda}=2\}}=1.
\end{eqnarray}
Therefore, 
\begin{eqnarray}\label{w1}
\lim_{\lambda \rightarrow 0}\mbox{Prob}\{T_{3,2}<T_{3,1}\}=1.
\end{eqnarray}\label{temp1}
Now, suppose that for $m<n$, the induction step, $\lim_{\lambda \rightarrow 0}\mbox{Prob}\{T_{m,2}<T_{m,1}\}=1.$ Then, equation \eqref{tem3} yields that
\begin{align}\label{tem31}
\lim_{\lambda \rightarrow 0}\frac{1}{\mbox{Prob}\{T_{n,2}<T_{n,1}\}}&=\lim_{\lambda \rightarrow 0}\frac{1-\mbox{Prob}\{Z_{n,\lambda}=n\}}{\mbox{Prob}\{Z_{n,\lambda}=2\}+\sum_{m=3}^{n-1}\mbox{Prob}\{Z_{n,\lambda}=m\}\mbox{Prob}\{T_{m,2}<T_{m,1}\}}\nonumber\\
&=\lim_{\lambda \rightarrow 0}\frac{\sum_{m=1}^{n-1}\mbox{Prob}\{Z_{n,\lambda}=m\}}{\mbox{Prob}\{Z_{n,\lambda}=2\}+\sum_{m=3}^{n-1}\mbox{Prob}\{Z_{n,\lambda}=m\}\mbox{Prob}\{T_{m,2}<T_{m,1}\}}\nonumber\\
&=\lim_{\lambda \rightarrow 0}\frac{\sum_{m=1}^{n-1}\mbox{Prob}\{Z_{n,\lambda}=m\}}{\mbox{Prob}\{Z_{n,\lambda}=2\}+\sum_{m=3}^{n-1}\mbox{Prob}\{Z_{n,\lambda}=m\}\left[\lim_{\lambda \rightarrow 0}\mbox{Prob}\{T_{m,2}<T_{m,1}\}\right]} \nonumber\\
&=\lim_{\lambda \rightarrow 0}\frac{\sum_{m=1}^{n-1}\mbox{Prob}\{Z_{n,\lambda}=m\}}{\mbox{Prob}\{Z_{n,\lambda}=2\}+\sum_{m=3}^{n-1}\mbox{Prob}\{Z_{n,\lambda}=m\}} \nonumber\\
&=\lim_{\lambda \rightarrow 0}\frac{\sum_{m=1}^{n-1}\mbox{Prob}\{Z_{n,\lambda}=m\}}{\sum_{m=2}^{n-1}\mbox{Prob}\{Z_{n,\lambda}=m\}} \nonumber\\
&=1+\lim_{\lambda \rightarrow 0}\frac{\mbox{Prob}\{Z_{n,\lambda}=1\}}{\sum_{m=2}^{n-1}\mbox{Prob}\{Z_{n,\lambda}=m\}} \nonumber\\
&=1,
\end{align}
where that last equality follows  because $\lim_{\lambda \rightarrow 0}\frac{\mbox{Prob}\{Z_{n,\lambda}=1\}}{\sum_{m=2}^{n-1}\mbox{Prob}\{Z_{n,\lambda}=m\}}=0$ since (by Sandwich theorem)
\begin{align*}
0\leq \lim_{\lambda \rightarrow 0}\frac{\mbox{Prob}\{Z_{n,\lambda}=1\}}{\sum_{m=2}^{n-1}\mbox{Prob}\{Z_{n,\lambda}=m\}} \leq \lim_{\lambda \rightarrow 0}\frac{\mbox{Prob}\{Z_{n,\lambda}=1\}}{\mbox{Prob}\{Z_{n,\lambda}=2\}}=0,
\end{align*}
where the last equality follows by \eqref{tem1}. \\
Finally, by \eqref{tem31}, we conclude that
\begin{eqnarray}\label{w11}
\lim_{\lambda \rightarrow 0}\mbox{Prob}\{T_{n,2}<T_{n,1}\}=1.
\end{eqnarray} 
Thus,  the proof of the second part of the theorem is complete. 
\\
\noindent{\textbf{Proof of III.}}:\\
Now we prove the last part of the theorem. By the definition  of $T_{n,m}$, we have
\begin{eqnarray}
\mathbb{E}[T_{n,m}]=\frac{\mbox{Prob}\{Z_{n,\lambda}\leq m\}+\sum_{k=m+1}^{n-1}\mbox{Prob}\{Z_{n,\lambda}=k\}\mathbb{E}[T_{k,m}]}{1-\mbox{Prob}\{Z_{n,\lambda}=n\}}.
\end{eqnarray} 
Thus,
\begin{eqnarray}\label{kk1}
\frac{\mathbb{E}[T_{n,2}]}{\mathbb{E}[T_{n,1}]}=\frac{\mbox{Prob}\{Z_{n,\lambda}=1\}+\mbox{Prob}\{Z_{n,\lambda}=2\}+\sum_{k=3}^{n-1}\mbox{Prob}\{Z_{n,\lambda}=k\}\mathbb{E}[T_{k,2}]}{\mbox{Prob}\{Z_{n,\lambda}=1\}+\sum_{k=2}^{n-1}\mbox{Prob}\{Z_{n,\lambda}=k\}\mathbb{E}[T_{k,1}]}.
\end{eqnarray} 
Now, by induction over $n$ we show that
\begin{eqnarray}
\lim_{\lambda \rightarrow 0}\frac{\mathbb{E}[T_{n,2}]}{\mathbb{E}[T_{n,1}]}=0.
\end{eqnarray}
Suppose that $n=3$, then

\begin{eqnarray}
\lim_{\lambda \rightarrow 0}\frac{\mathbb{E}[T_{3,2}]}{\mathbb{E}[T_{3,1}]}&=&\lim_{\lambda \rightarrow 0}\frac{\mbox{Prob}\{Z_{3,\lambda}=1\}+\mbox{Prob}\{Z_{3,\lambda}=2\}}{\mbox{Prob}\{Z_{3,\lambda}=1\}+\mbox{Prob}\{Z_{3,\lambda}=2\}\mathbb{E}[T_{2,1}]}\nonumber \\
&=&0.
\end{eqnarray} 

The last equality follows because 
$$\lim_{\lambda \rightarrow 0}\frac{\mbox{Prob}\{Z_{3,\lambda}=1\}}{\mbox{Prob}\{Z_{3,\lambda}=2\}}=0$$ by Eq. \eqref{tem1} and $$\lim_{\lambda \rightarrow 0}\mathbb{E}[T_{2,1}]=\infty$$ by the first part of the theorem.

Now, suppose that for $k<n$, the induction step,
\begin{eqnarray}\label{hasan}
\lim_{\lambda \rightarrow 0}\frac{\mathbb{E}[T_{k,2}]}{\mathbb{E}[T_{k,1}]}=0.
\end{eqnarray} 
Before completing the proof,  first we have the following Lemma that is useful in the sequel.
\begin{Lemma}\label{lem22}
If $0 \leq a_i$ and $ 0 < b_i$ for $i=1,2\cdots, n$, then $\frac{\sum_{i=1}^n a_i}{\sum_{i=1}^n b_i}\leq \sum_{i=1}^n\frac{a_i}{b_i}$.
\end{Lemma}
\begin{proof}
The proof is immediate since
\begin{align*}
\frac{\sum_{i=1}^n a_i}{\sum_{i=1}^n b_i}=\sum_{i=1}^n\frac{a_i}{\sum_{i=1}^n b_i}\leq \sum_{i=1}^n\frac{a_i}{b_i}.
\end{align*}
\end{proof}
Now, in the following, using the induction step we show the statement is valid for $k=n$ as well. We do this  by employing the Sandwich theorem as follows.
\begin{eqnarray}
0 \leq \lim_{\lambda \rightarrow 0} \frac{\mathbb{E}[T_{n,2}]}{\mathbb{E}[T_{n,1}]}&\stackrel{(a)}{=}& \lim_{\lambda \rightarrow 0} \frac{\mbox{Prob}\{Z_{n,\lambda}=1\}+\mbox{Prob}\{Z_{n,\lambda}=2\}+\sum_{k=3}^{n-1}\mbox{Prob}\{Z_{n,\lambda}=k\}\mathbb{E}[T_{k,2}]}{\mbox{Prob}\{Z_{n,\lambda}=1\}+\sum_{k=2}^{n-1}\mbox{Prob}\{Z_{n,\lambda}=k\}\mathbb{E}[T_{k,1}]} \nonumber\\
&\stackrel{(b)}{\leq}& \lim_{\lambda \rightarrow 0} \frac{\mbox{Prob}\{Z_{n,\lambda}=2\}+\sum_{k=3}^{n-1}\mbox{Prob}\{Z_{n,\lambda}=k\}\mathbb{E}[T_{k,2}]}{\sum_{k=2}^{n-1}\mbox{Prob}\{Z_{n,\lambda}=k\}\mathbb{E}[T_{k,1}]} \nonumber\\
& \stackrel{(c)}{\leq} & \lim_{\lambda \rightarrow 0} \left[ \frac{\mbox{Prob}\{Z_{n,\lambda}=2\}}{\mbox{Prob}\{Z_{n,\lambda}=2\}\mathbb{E}[T_{k,2}]}+ \sum_{k=3}^{n-1}\frac{\mbox{Prob}\{Z_{n,\lambda}=k\}\mathbb{E}[T_{k,2}]}{\mbox{Prob}\{Z_{n,\lambda}=k\}\mathbb{E}[T_{k,1}]} \right]\nonumber\\
&=&  \lim_{\lambda \rightarrow 0}  \frac{1}{\mathbb{E}[T_{k,2}]} + \sum_{k=3}^{n-1} \left[  \lim_{\lambda \rightarrow 0} \frac{\mathbb{E}[T_{k,2}]}{\mathbb{E}[T_{k,1}]} \right] \nonumber\\
&\stackrel{(d)}{=}&0.
\end{eqnarray} 
where (a) is followed by \eqref{kk1}, (b) is correct since 
$$\lim_{\lambda \rightarrow 0} \frac{\mbox{Prob}\{Z_{n,\lambda}=1\}}{\mbox{Prob}\{Z_{n,\lambda}=1\}+\sum_{k=2}^{n-1}\mbox{Prob}\{Z_{n,\lambda}=k\}\mathbb{E}[T_{k,1}]}=0, 
$$
(c) is correct by Lemma \ref{lem22}, and  finally (d) is followed by  \eqref{hasan}. By the above equality the proof of the third part of the theorem is complete.
\end{proof}

\end{document}